\documentclass[journal]{IEEEtran}
\usepackage{amsmath,graphicx}
\usepackage{epstopdf}
\usepackage{color}
\usepackage{wrapfig}
\usepackage{lipsum}
\usepackage{float}
\usepackage{colortbl,booktabs}
\usepackage{mathtools}
\usepackage[compress]{cite}
\definecolor{myRed}{rgb}{0.8, 0.2, 0.2}
\definecolor{myYellow}{rgb}{0.2,0.2,0.8}

\newcommand{\figref}[1]{Fig.~\ref{#1}}

\newcommand{\secref}[1]{Sec.~\ref{#1}}
\newcommand{\eqnref}[1]{(\ref{#1})}

\usepackage{amssymb}
\newcounter{MYtempeqncnt}
\usepackage{multirow}
\usepackage{float}
 \usepackage{bm}
\usepackage{subfigure}
\usepackage{multicol}
\usepackage{algorithmic}
\usepackage{algorithm}
 \usepackage{mathrsfs}
\usepackage{bm}
\usepackage{amsthm}
\theoremstyle{definition}
\newtheorem{defn}{Definition}
\newtheorem{prop}{Proposition}
\theoremstyle{remark}
\newtheorem{remark}{Remark}
\usepackage{cite}
\usepackage{stfloats}
\ifCLASSINFOpdf
\else
\fi
\begin{document}

\title{On the Performance of Splitting Receiver with Joint Coherent and Non-Coherent Processing}
\author{Yanyan Wang, Wanchun Liu, \emph{Member, IEEE}, Xiangyun Zhou, \emph{Senior Member, IEEE}, \\ and Guanghui Liu, \emph{Senior Member, IEEE}
}

\maketitle

\IEEEpeerreviewmaketitle
\begin{abstract}
\let\thefootnote\relax\footnote{
	Y. Wang is with School of Information and Communication Engineering, University of Electronic Science and Technology of China, Chengdu 611731, China  (email: yywang@std.uestc.edu.cn), and was also a visiting student at Research School of Electrical, Energy and Materials Engineering, The Australian National University, Australia.
	W. Liu is with School of Electrical and Information Engineering, The University of
	Sydney, Australia (email: wanchun.liu@sydney.edu.au). Her work was supported by the Australian Research Council’s Australian Laureate Fellowships Scheme under Project FL160100032.
	X. Zhou is with Research School of Electrical, Energy and Materials Engineering, The Australian National University, Australia  (email: xiangyun.zhou@anu.edu.au).
	G. Liu is with School of Information and Communication Engineering, University of Electronic Science and Technology of China, Chengdu 611731, China  (email: guanghuiliu@uestc.edu.cn).	
}In this paper, we revisit a recently proposed receiver design, named the splitting receiver, which jointly uses coherent and non-coherent processing for signal detection. By considering an improved signal model for the splitting receiver as compared to the original study in the literature, we conduct a performance analysis on the achievable data rate under Gaussian signaling and obtain a fundamentally different result on the performance gain of the splitting receiver over traditional receiver designs that use either coherent or non-coherent processing alone. Specifically, the original study ignored the antenna noise and concluded on a 50\% gain in achievable data rate in the high signal-to-noise ratio (SNR) regime. In contrast, we include the antenna noise in the signal model and show that the splitting receiver improves the achievable data rate by a constant gap in the high SNR regime. This represents an important correction of the theoretical understanding on the performance of the splitting receiver. In addition, we examine the maximum-likelihood detection and derive a low-complexity detection rule for the splitting receiver for practical modulation schemes. Our numerical results give further insights into the conditions under which the splitting receiver achieves significant gains in terms of either achievable data rate or detection error probability.
\end{abstract}

\begin{IEEEkeywords}
Splitting receiver, wireless receiver, coherent detection, power detection, maximum-likelihood detection.
\end{IEEEkeywords}
\section{Introduction}
\subsection{Background and Prior Work}

As described by ``Cooper's Law", the data rate of wireless communications has approximately doubled about every two years since the advent of the cellular phone. The everlasting demand for higher data rate has attracted a significant amount of research into several promising technologies~\cite{PetarMag}. Many of the cutting-edge communication technologies make use of the increasing number of multiple antennas for transmission and/or reception, e.g., massive multiple-input multiple-output~(MIMO)~\cite{larsson2014massive}. However, the basic design principles of a radio-frequency (RF) digital receiver behind each antenna have stayed virtually unchanged throughout the evolution of wireless digital communications.

Roughly speaking, there are two types of receivers commonly used or considered in wireless communication systems: coherent and non-coherent receivers.
A coherent receiver is based on {coherent detection} (CD)~\cite{proakis2007digital}, where the received RF-band signal is converted to a baseband signal with in-phase (I) and quadrature (Q) components by using a down-conversion circuit, which is then digitized through an analog-to-digital converter (ADC).
The coherent receiver design is rather mature and has been adopted in most of the wireless communication standards, and the current research mostly focuses on two directions: I/Q imbalance compensation due to the hardware imperfection~\cite{IQ} and low-power design for both cellular and wireless sensor networks~\cite{lowpower}. For the latter, low-resolution or even one-bit ADCs have been considered for massive MIMO systems~\cite{Mo,Li}, since high-resolution ADCs are power-hungry for portable devices.

Among different non-coherent receiver designs, the power detection (PD) based receiver~\cite{Chowdhury2016Scaling,Jing2016Design} converts the RF-band signal directly to a direct current (DC) signal representing the power or intensity, by using a rectifier circuit, which is then digitized using an ADC.
Such a PD-based non-coherent receiver typically suffers from certain performance losses compared to the coherent receiver. Nevertheless, it has been proved that the achievable rate of the PD-based non-coherent receiver exhibits the same scaling law as the CD system in the limit of a large number of receiving antennas~\cite{Chowdhury2016Scaling}.
More importantly, the PD-based non-coherent receiver has lower power consumption than the CD receiver in general, since the rectifier circuit of a PD receiver consists of passive diodes while the RF-to-baseband conversion circuit of a CD receiver has active components including oscillators and mixers. Thus, the PD-based non-coherent receiver can provide power-efficient solutions for massive receiver arrays~\cite{Jing2016Design} and ultra-low power wireless sensors~\cite{liu2019next}.

Recently, a new receiver architecture, named the splitting receiver, was proposed in~\cite{Liu2017A}, which marks another major innovation in the basic design principles of wireless receiver. In the nutshell, the splitting receiver utilizes both CD and PD for information detection. To be specific, by using a passive RF power splitter, the received signal at the splitting receiver is split into two streams, which are then processed by a CD circuit and a PD circuit, separately. Then, the processed two-stream signals are jointly utilized for information detection. Thus, the splitting receiver introduces an interesting signal-processing method of joint coherent and non-coherent processing. Using a simplified analytical model for the receiver, the work in~\cite{Liu2017A} showed in the high  signal-to-noise ratio (SNR) regime that the splitting receiver is able to achieve 50\% higher rate than either the traditional CD receiver or the PD receiver. In addition, when applying the standard $M$-quadrature amplitude modulation (QAM) for information transmission, the splitting receiver was shown to reduce the symbol error rate (SER) by a factor of $\sqrt{M}-1$ as compared to the traditional receivers in the high SNR regime.

\subsection{Motivation and Novel Contributions}
\begin{figure}[t]
	\centering
	\includegraphics[width=1.0 \linewidth]{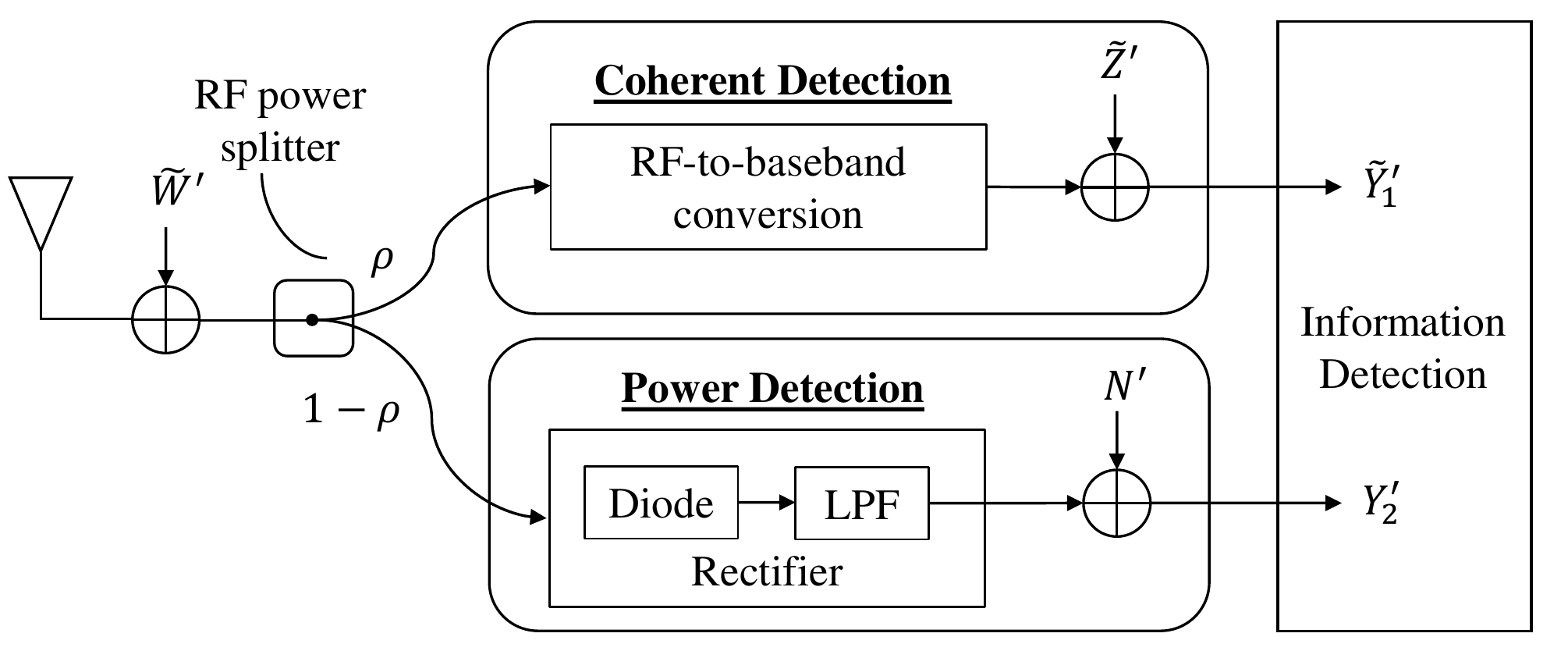}
	\caption{The splitting receiver architecture with antenna noise and processing noises.}
	\label{fig:Fig1sysmod}
\end{figure}

While the new receiver architecture proposed in~\cite{Liu2017A} opens up an exciting research direction for the use of such a receiver in various wireless systems and networks, there still needs significant effort in establishing a better understanding on the fundamental performance limits of the splitting receiver, even for the most basic case of a single-antenna receiver. Taking a closer look at the receiver model considered in~\cite{Liu2017A}, we see that it only considered the processing noises in the CD and PD circuits while ignored the antenna noise. In other words, for the splitting receiver model shown in ~\figref{fig:Fig1sysmod}, the work in~\cite{Liu2017A} only considered the processing noise of the CD circuit $\tilde{Z}'$ and the processing noise of the PD circuit $N'$ but ignored the antenna noise $\tilde{W}'$. Hence, the performance characterization of the splitting receiver obtained in~\cite{Liu2017A} is only valid for the antenna-noise-free case. One might expect that the results in the antenna-noise-free case should still be accurate even when there is antenna noise as long as it is much smaller than the processing noises. However, we argue that the antenna noise (before splitting the signal) and the processing noises (after the signal is split into two streams) have fundamentally different impacts on the signal quality for the specific receiver architecture. This motivates us to revisit the performance of the splitting receiver by taking into account the antenna noise.

In this paper, we aim to establish the performance limits of a single-antenna splitting receiver considering both the antenna noise and processing noises. We analyze the performance in terms of achievable mutual information (i.e., data rate) as well as the SER of signal detection and obtain the following novel results:

\begin{itemize}
  \item We analyze the achievable mutual information of the splitting receiver under Gaussian signaling and derive a closed-form approximation of the mutual information which is asymptotically tight at high SNR. From this result, we characterize the mutual information performance gain of the splitting receiver as compared with the traditional CD and PD receivers. Specifically, we show that the splitting receiver improves the achievable mutual information by a constant gap at high SNR. Therefore, the percentage gain in mutual information reduces asymptotically to zero as SNR approaches infinity. This result is fundamentally different from the result obtained in~\cite{Liu2017A} where an asymptotic percentage gain of 50\% was concluded under the antenna-noise-free assumption. Therefore, the result in this paper fundamentally changes, and more importantly, corrects the state-of-the-art understanding on the mutual information performance of the splitting receiver.

  \item Apart from the analytical results obtained in the asymptotic high SNR regime, we also numerically study the mutual information gain of the splitting receiver for a large range of finite SNR values. Our results show that the splitting receiver can achieve significantly higher mutual information as compared to the traditional CD and PD receivers. For example, the observed percentage gain goes up to 44.2\% among the cases we examined. The results highlight the advantage of using the splitting receiver for improving data rate at practical SNR values.

  \item We also examine the maximum-likelihood (ML) detection rule for the splitting receiver for practical modulations. The optimal ML detector is found to have high computational complexity. To facilitate practical implementation, we derive a low-complexity detection rule, which is asymptotically optimal at high SNR. The SER performance of QAM modulation is then numerically studied and the results again demonstrate the advantage of using the splitting receiver as opposed to the traditional CD and PD receivers. For example, the numerical results obtained for 64-QAM show that, to achieve the same target SER of $10^{-2}$, the traditional CD receiver requires 32\% more transmit power than the splitting receiver, under certain noise conditions.
\end{itemize}

\subsection{Paper Organization and Notation}

The remainder of the paper is organized as follows: Section~II describes the signal model for the splitting receiver. The mutual information performance analysis is carried out in Section~III. The ML and low-complexity detection rules, as well as the SER performance for practical modulations are presented in Section~IV. Finally, Section~V concludes this work and points out future research directions.

\textit{Notation}: $\tilde{\cdot}$ represents a complex number. $|\cdot|$ is the absolute-value norm of a complex number.  $\mathcal{H}
(\cdot)$, $\mathcal{H}(\cdot, \cdot)$, $\mathcal{H}(\cdot|\cdot)$
are the differential entropy, joint and conditional differential
entropy, respectively. $\mathcal{I} (\cdot; \cdot)$ denotes the mutual information. $(\cdot)_r$ and $(\cdot)_i$ denote the real part and imaginary part of a complex number, respectively. In addition, $\mathcal{N}(m,\sigma^2)$ and $\mathcal{CN}(m,\sigma^2)$ denote the real-valued and complex-valued Gaussian distribution with mean $m$ and variance $\sigma^2$. $\mathbb{E}(\cdot)$ and $\textrm{Var}(\cdot)$ denote the expectation and variance of a random variable, respectively.

\section{System Model}
We consider a baseline single-antenna point-to-point system and focus our attention at the receiver side. The transmitted signal goes through a channel with coefficient $\tilde{h}\triangleq|\tilde{h}|e^{j\phi}$ and reaches the receiver. We assume that the channel coefficient is constant during the communication and known at the receiver. \figref{fig:Fig1sysmod} shows the splitting receiver architecture. The received RF signal at the receiving antenna is first corrupted by the antenna noise $\tilde{W}^{'}$. Then, the received signal is divided into two streams by a passive power splitter, e.g., a Wilkinson power divider~\cite{Wilkinsons}, which is theoretically lossless and has ultra-low power loss in practice. Thus, we assume that the power splitter is ideal at the receiver~\cite{zhang2013mimo,wu2010analytical}. The two streams are processed by a conventional CD circuit and a PD circuit, respectively. The CD circuit applies RF-to-baseband conversion and digitizes the baseband signal through an ADC. The rectifier-based PD circuit converts the RF signal into a DC signal. The processing noises for the CD and PD circuits are denoted as $\tilde{Z}^{'}$ and $N^{'}$, respectively.

In this paper, we model the antenna noise and processing noises as zero-mean Gaussian noises. The antenna noise and processing noise in the CD circuit are commonly modelled using Gaussian distributions. On the other hand, the Gaussian noise modelling for the PD circuit is a largely simplified model. This is because that the Gaussian modeling of the processing noise in the PD circuit can cause the issue of having a negative output signal, while the output signal should be strictly non-negative. However, this issue is also negligible if the operating SNR is moderately large, which is indeed the regime that this work focuses on. In addition, prior experimental studies on the properties of the main types of noise in the rectifier circuit showed that Gaussian distribution is a reasonable model~\cite{PhysRevLett}. Gaussian noise modeling is also a common analytical approach for investigating the performance of non-coherent receivers (having non-negative output signals)~\cite{zhou2013wireless,lapidoth2009capacity,infrared1997}.

In practice, the antenna noise consists of ambient RF signals (i.e., interference signals) within the same spectrum of the target signal and the thermal noise induced by the antenna circuit \cite{gu2006rf}.
In an interference-limited scenario~\cite{lin2013achieving, yang2018dense, perez2018signal}, the interference has the same level as the useful signal, and hence,
the antenna noise is dominated by the interference and can be much stronger than the processing noises.
In an interference-free scenario (or more practically a noise-limited scenario), the antenna noise is approximately at the thermal noise level, which is much smaller than the processing noises. More specifically, the thermal noise power is given as $kTB$~\cite{razavi1998rf}, where $k=1.38\times10^{-23}$~J/K denotes the Boltzmann constant, $T$ is the absolute temperature and $B$ is the bandwidth of interest. Taking a signal bandwidth of 10~MHz and temperature of 290 K for an example, the thermal noise power is approximately $-104$~dBm. Taking a commercial CD circuit, AD9870~\cite{hendriks2001high} as an example, when the receiving signal power is $-40$~dBm, the power of the processing noise is about two orders of magnitude times higher than the thermal noise power. For the PD circuit~\cite{loy1999understanding, clerckx2018fundamentals}, the rectifier is composed of one or multiple diodes followed by a low-pass filter with a load. Taking the Schottky diode reported in~\cite{flicker} as an example, when the receiving signal power is $-40$~dBm, the noise generated by the Schottky diode is about $-85$~dBm, which can be used as a rough indication of rectifier noise (also taking into account how many Schottky diodes are used in the rectifier circuit). Hence, the rectifier noise power can easily be at least two orders of magnitude times higher than the thermal noise power. In short, depending on whether it is an interference-limited or a noise-limited scenario, the relative strengths of the antenna noise and processing noises can vary significantly.

In the original work of splitting-receiver design~\cite{Liu2017A}, the receiver performance was analyzed assuming no antenna noise, i.e., $\tilde{W}^{'}=0$. As we will see in the sequel of the paper, the omission of the antenna noise can lead to fundamentally different understanding on the receiver performance. Therefore, the previous (antenna-noise-free) results in~\cite{Liu2017A} do not give the correct understanding of the receiver performance even when the antenna noise is much weaker than the processing noises.

Taking the antenna noise into account, the post-processing baseband signals of the CD and PD circuits are given by
\begin{equation}
\label{equ:CDreceiverc}
{{{\tilde{Y}}}^{'}_1} = \sqrt{\rho}(\sqrt{P}\tilde{h}{{\tilde{X}}}+{{\tilde{W}^{'}}})+{{\tilde{Z}^{'}}},
\end{equation}
\begin{equation}
\label{equ:PDreceiverc}
{{{Y}}^{'}_2} = \eta (1-\rho)|\sqrt{P}\tilde{h}{{\tilde{X}}}+{{\tilde{W}^{'}}}|^2+{{N}}^{'},
\end{equation}
where $\rho \in [0,1]$ is the power splitting ratio. When $\rho=0$, the splitting receiver is degraded to the non-coherent receiver. When $\rho=1$, the splitting receiver is degraded to the coherent receiver. Also, $\eta\in [0,1]$ denotes the conversion efficiency of the rectifier-based PD circuit. ${{\tilde{X}}}$ is the transmitted signal with normalized variance and $P$ is the average transmit power of the signal.

For the ease of analysis and without loss of generality, we apply some simple manipulation on the baseband signals as ${{{\tilde{Y}}}_1} = e^{-j\phi}{{{\tilde{Y}}}^{'}_1}$ and ${{{Y}}_2}={{{Y}}^{'}_2}/\eta$. Hence, the equivalent baseband signals are given by
\begin{equation}
\label{equ:CDreceiver}
{{{\tilde{Y}}}_1} = \sqrt{\rho}(\sqrt{P}|\tilde{h}|{{\tilde{X}}}+{{\tilde{W}}})+{{\tilde{Z}}},
\end{equation}
\begin{equation}
\label{equ:PDreceiverch}
{{{Y}}_2} = (1-\rho)\big|\sqrt{P}{|\tilde{h}|}{{\tilde{X}}}+{{\tilde{W}}}\big|^2+{{N}},
\end{equation}
where ${\tilde{W}}\triangleq e^{-j\phi}{{\tilde{W}}^{'}}$, $\tilde{Z}\triangleq e^{-j\phi}{{\tilde{Z}^{'}}}$, and ${{N}} \triangleq {{N}}^{'}/ \eta$.

In the remainder of the paper, we use \eqnref{equ:CDreceiver} and \eqnref{equ:PDreceiverch} as the received baseband signals for further analysis. With slight abuse of language, we call ${\tilde{W}}$, $\tilde{Z}$, and ${{N}}$ as the antenna noise, conversion noise, and rectifier noise, respectively. Note that these noises still follow zero-mean Gaussian distributions. We denote their variances by $\sigma_{\textrm{A}}^2$, $\sigma_{\textrm{cov}}^2$, and $\sigma_{\textrm{rec}}^2$, respectively. Also, while the magnitude of the constant channel coefficient $|\tilde{h}|$ is kept generic in all analytical results, we will set it as $|\tilde{h}|=1$ in all numerical results for simplicity. We will also set the rectifier's conversion efficiency as $\eta = 1$ in all numerical results for simplicity.

\section{Mutual Information Performance Analysis}
In this section, we analyze the mutual information achieved by using the splitting receiver. This gives insights into the fundamental limit of the achievable rate for such a receiver architecture. In particular, we compare the maximum mutual information achieved by the splitting receiver with the optimal splitting ratio (i.e., optimal $\rho$) against the mutual information achieved by the traditional CD receiver (effectively setting $\rho=1$) or the traditional PD receiver (effectively setting $\rho=0$), in order to find out how much performance gain there is by using the splitting receiver architecture.

\subsection{Mutual Information of the Splitting Receiver}
From an information-theoretic perspective, \eqnref{equ:CDreceiver} and \eqnref{equ:PDreceiverch} jointly describe the input-output relationship of the equivalent baseband channel, where the input is $\sqrt{P}|\tilde{h}|{{\tilde{X}}}$ and the output is $({{{\tilde{Y}}}_1}, {{{Y}}_2})$. The mutual information between the input and output is expressed as
\begin{align}
\label{equ:MISPreceiver}
&\mathcal{I}(\sqrt{P}|\tilde{h}|{{\tilde{X}}};{{{\tilde{Y}}}_1},{{{Y}}_2})\notag\\
& =\mathcal{H}({{{\tilde{Y}}}_1},{{{Y}}_2})-\mathcal{H}({{{\tilde{Y}}}_1},{{{Y}}_2}\big|\sqrt{P}|\tilde{h}|{{\tilde{X}}})\notag\\
& =\mathcal{H}({{{\tilde{Y}}}_1},{{{Y}}_2})-\int_{{{\tilde{X}}}}f_{{{\tilde{X}}}}(\tilde{x})
\mathcal{H}({{{\tilde{Y}}}_1},{{{Y}}_2}\big|\sqrt{P}|\tilde{h}|\tilde{x})\,\mathrm{d}\tilde{x}\notag\\
&=-\int_{{{{Y}}_2}}\int_{{{{\tilde{Y}}}_1}}f_{{{{\tilde{Y}}}_1},{{{Y}}_2}}({{{\tilde{y}}}_1},{{{y}}_2})\log_2\left(f_{{{{\tilde{Y}}}_1},{{{Y}}_2}}({{{\tilde{y}}}_1},{{{y}}_2})\right) \,\mathrm{d}{{{\tilde{y}}}_1}\mathrm{d}{{{y}}_2}\notag\\
&\,\,\,\,\,\,\,\,+\int_{{{\tilde{X}}}}\int_{{{\tilde{Y}_1}}}\int_{{{{Y_2}}}}f_{{{\tilde{X}}}}(\tilde{x})f_{{{\tilde{Y}}_1},{{{Y_2}}}}(\tilde{y}_1,{y_2}|\tilde{x})\notag\\
&\,\,\,\,\,\,\,\,\,\,\,\,\,\,\,\log_2 \left(f_{{{\tilde{Y}_1}},{{{Y_2}}}}(\tilde{y}_1,{y_2}|\tilde{x})\right)\,\mathrm{d}{y_2}\mathrm{d}\tilde{y}_1\mathrm{d}\tilde{x}.
\end{align}

Note that the integration over the complex-valued $\tilde{X}$ or $\tilde{Y}_1$ is taken over a two-dimensional space (i.e., two one-dimensional integrals) and the integration over the real-valued ${Y}_2$ is one dimensional. Furthermore, evaluating the mutual information requires a number of joint and conditional probability density functions (PDFs). Specifically, the joint PDF of $({{{\tilde{Y}}}_1},{{{Y}}_2})$ is given as
\begin{align}
\label{equ:pdfy1y2}
&f_{{{{\tilde{Y}}}_1},{{{Y}}_2}}({{{\tilde{y}}}_1},{{{y}}_2})\notag\\
&=\int_{\tilde{X}}\int_{\tilde{W}}f_{{{{\tilde{Y}}}_1},{{{Y}}_2}}({{{\tilde{y}}}_1},{{{y}}_2}|\tilde{x},\tilde{w})
f_{{{\tilde{X}}}}(\tilde{x})f_{{{\tilde{W}}}}(\tilde{w})\,\mathrm{d}{\tilde{w}}\mathrm{d}{\tilde{x}}\notag\\
&=\int_{\tilde{X}}\int_{\tilde{W}}f_{{{{\tilde{Y}}}_1}}({{{\tilde{y}}}_1}|\tilde{x},\tilde{w})f_{{{{Y}}_2}}({{{y}}_2}|\tilde{x},\tilde{w})
f_{{{\tilde{X}}}}(\tilde{x})f_{{{\tilde{W}}}}(\tilde{w})\,\mathrm{d}{\tilde{w}}\mathrm{d}{\tilde{x}},
\end{align}
and the conditional joint PDF $f_{{{\tilde{Y}_1}},{{{Y_2}}}}(\tilde{y}_1,{y_2}|\tilde{x})$ is given as
\begin{align}
\label{equ:pdfy1y2x}
f_{{{\tilde{Y_1}}},{{{Y_2}}}}(\tilde{y}_1,{y_2}|\tilde{x})
&=\int_{\tilde{W}}f_{{{\tilde{W}}}}(\tilde{w})f_{{{{\tilde{Y}}}_1},{{{Y}}_2}}({{{\tilde{y}}}_1},{{{y}}_2}|\tilde{x},\tilde{w})\,\mathrm{d}{\tilde{w}}\notag\\
&=\!\int_{\tilde{W}}\!f_{{{{\tilde{Y}}}_1}}({{{\tilde{y}}}_1}|\tilde{x},\tilde{w})f_{{{{Y}}_2}}({{{y}}_2}|\tilde{x},\tilde{w})f_{{{\tilde{W}}}}(\tilde{w})\,\mathrm{d}{\tilde{w}},
\end{align}
where $f_{{{\tilde{X}}}}(\tilde{x})$ is the PDF of the normalized input signal, $f_{{{\tilde{W}}}}(\tilde{w})$ follows $\mathcal{CN}\big(0,\sigma_{\textrm{A}}^2\big)$, and the conditional PDFs $f_{{{{\tilde{Y}}}_1}}({{{\tilde{y}}}_1}|\tilde{x},\tilde{w})$ and $f_{{{{Y}}_2}}({{{y}}_2}|\tilde{x},\tilde{w})$ are $\mathcal{CN}\big(\sqrt{\rho}(\sqrt{P}|\tilde{h}|\tilde{x}+\tilde{w}),\sigma_{\textrm{cov}}^2\big)$ and $\mathcal{N}\big( (1-\rho)\big|\sqrt{P}{|\tilde{h}|}{{\tilde{x}}}+{{\tilde{w}}}\big|^2,\sigma_{\textrm{rec}}^2\big)$, respectively.

It is not hard to see that both terms in the mutual information expression in \eqnref{equ:MISPreceiver} eventually have seven one-dimensional integrals, which is extremely cumbersome to compute. Also, it is extremely challenging to identify the optimal input distribution $f_{{{\tilde{X}}}}(\tilde{x})$ that maximizes the mutual information. For tractability and ease of obtaining analytical insights, we assume that the input signal follows a normalized Gaussian distribution, i.e., $\tilde{X}\sim \mathcal{CN}(0,1)$. It is well-known that this is the optimal input distribution for the traditional CD receiver. In what follows, we proceed our study by considering three cases with different power splitting ratios.

1) When $\rho=1$, the splitting receiver is degraded to the traditional CD receiver. The mutual information is given as
\begin{align}
\label{equ:MICDreceiver}
\mathcal{I}(\sqrt{P}|\tilde{h}|{{\tilde{X}}};{{{\tilde{Y}}}_1},{{{Y}}_2})& = \mathcal{H}({{{\tilde{Y}}}_1})-\mathcal{H}({{\tilde{W}}}+{{\tilde{Z}}})\notag\\
& =\log_2\left(1+\frac{P|\tilde{h}|^2}{\sigma_{\textrm{A}}^2+\sigma_{\textrm{cov}}^2}\right).
\end{align}

2) When $\rho=0$, the splitting receiver is degraded to the PD-based non-coherent receiver. The mutual information is given as
\begin{align}
\label{equ:MIPDreceiver}
&\mathcal{I}(\sqrt{P}|\tilde{h}|{{\tilde{X}}};{{{\tilde{Y}}}_1},{{{Y}}_2}) \notag\\
&= \mathcal{H}({{{{Y}}}_2})-\mathcal{H}({{{{Y}}}_2}\big |\sqrt{P} |\tilde{h}|{{\tilde{X}}})\notag\\
& =\mathcal{H}({{{{Y}}}_2})-\mathcal{H}\big(\big|\sqrt{P}|\tilde{h}|{{\tilde{X}}}+{{\tilde{W}}}\big|^2+{{N}}\big|\sqrt{P}|\tilde{h}|{{\tilde{X}}}\big)\notag\\
&=\mathcal{H}({{{{Y}}}_2})-\int_{{{\tilde{X}}}}f_{{{\tilde{X}}}}(\tilde{x}) \mathcal{H}\big(\big|\sqrt{P}|\tilde{h}|\tilde{x}+{{\tilde{W}}}\big|^2+{{N}}\big|\sqrt{P}|\tilde{h}|\tilde{x}\big)\,\mathrm{d}\tilde{x} \notag\\
&=-\int_{{{{Y}}}_2}{f_{{{{Y}}_2}}(y_2)\log_2\big(f_{{{{Y}}_2}}(y_2)\big)\,\mathrm{d}{y_2}}\notag\\
&\,\,\,\,\,\,\,+\int_{{{\tilde{X}}}}\int_{{{{R}}}_c}{f_{{{{R}}_c}}(r_c)\log_2\big(f_{{{{R}}_c}}(r_c)\big)f_{{{\tilde{X}}}}(\tilde{x})\,\mathrm{d}{r_c}\mathrm{d}\tilde{x}},
\end{align}
where we have defined two new random variables, namely ${{{{R}}_c}} \triangleq {{{{R}}_n}} + {{N}}$ and ${{{R}}_{n}}\triangleq \big|\sqrt{P}|\tilde{h}|{{\tilde{x}}}+{{\tilde{W}}}\big|^2$. The random variable $Y_2$ follows an exponential modified Gaussian distribution \cite{haney2011practical}, \cite{grushka1972characterization}:
\begin{align}
\label{equ:fy2pdf}
f_{{{{Y}}_2}}(y_2)
=&\frac{1}{2(P|\tilde{h}|^2+\sigma_{\textrm{A}}^2)}{\operatorname{erfc}}\left(\frac{\frac{\sigma_{\textrm{rec}}^2}{P|\tilde{h}|^2+\sigma_{\textrm{A}}^2}-y_2}{\sqrt{2}\sigma_{\textrm{rec}}}\right)\notag\\
&\exp\left(\frac{1}{2(P|\tilde{h}|^2+\sigma_{\textrm{A}}^2)}\biggl(\frac{\sigma_{\textrm{rec}}^2}{P|\tilde{h}|^2+\sigma_{\textrm{A}}^2}-2y_2\biggr)\right),
\end{align}
where $\operatorname{erfc}(\cdot)$  is the complementary error function. In addition, for a given ${{\tilde{x}}}$, the random variable ${{{R}}_{n}}$ follows a noncentral chi-squared distribution \cite{siegel1979noncentral}:
\begin{equation}
\label{equ:fr2pdf}
\begin{aligned}
f_{{{{R}}_n}}(r_n)&= \frac{1}{2\sigma_{\textrm{s}}^2}\exp\left(-\frac{r_n+\lambda}{2\sigma_{\textrm{s}}^2}\right)\operatorname{I}_0\left(\frac{\sqrt{r_n\lambda}}{\sigma_{\textrm{s}}^2}\right),
\end{aligned}
\end{equation}
where $\sigma_{\textrm{s}}^2=\frac{\sigma_{\textrm{A}}^2}{2}$, $\lambda=P|\tilde{h}|^2|{\tilde{x}}|^2$, and $\operatorname{I}_0(\cdot)$ denotes the zeroth-order modified Bessel function of the first kind. Also knowing ${{N}}\sim \mathcal{N}(0,\sigma_{\textrm{rec}}^2)$, the distribution of ${{{{R}}_c}}$ can be obtained using the convolution of $f_{{{{R}}_n}}(r_n)$ and $f_N(n)$ as
\begin{align}
\label{equ:fyCpdf}
f_{{{{R}}_c}}(r_c)&= \int_{-\infty}^{+\infty} f_{{{{R}}_n}}(r_n)f_N({r_c}-{r_n})\,\mathrm{d}{r_n}\notag\\ &=\int_{-\infty}^{+\infty}\frac{1}{2\sqrt{2\pi}\sigma_{\textrm{s}}^2\sigma_{\textrm{rec}}}
\operatorname{I}_0\left(\frac{\sqrt{r_n\lambda}}{\sigma_{\textrm{s}}^2}\right)\notag\\
&\,\,\,\,\,\,\exp\left(\!-\frac{\sigma_{\textrm{rec}}^2(r_n\!+\!\lambda)+\sigma_{\textrm{s}}^2({r_c}\!-\!{r_n})^2}{2\sigma_{\textrm{s}}^2\sigma_{\textrm{rec}}^2}\right)\,\mathrm{d}r_n.
\end{align}

Based on \eqnref{equ:fy2pdf}, \eqnref{equ:fyCpdf} and $\tilde{X}\sim \mathcal{CN}(0,1)$, the mutual information in \eqnref{equ:MIPDreceiver} can be obtained.

3) When $\rho \in (0,1)$, due to the complicated form of \eqnref{equ:MISPreceiver}, it is extremely difficult to directly calculate the mutual information.
Nevertheless, the following proposition gives an accurate approximation of the mutual information in the high SNR regime\footnote{Note that the result in Proposition~\ref{proposition:prop1} is valid for $\rho \in (0,1)$ which does not include the two end values of $\rho = 0$ and $\rho = 1$.}.

\begin{prop}
The achievable mutual information of the splitting receiver with $\rho\in (0,1)$ can be approximated as
\begin{equation}
\label{equ:MIhighSNR}
\begin{aligned}
\mathcal{I}(\sqrt{P}|\tilde{h}|{{\tilde{X}}};{{{\tilde{Y}}}_1},{{{Y}}_2})\approx\log_2\left(\frac{\rho(P|\tilde{h}|^2+\sigma_{\textrm{{A}}}^2)}{\rho\sigma_{\textrm{{A}}}^2+\sigma_{\textrm{{cov}}}^2}\right)
+\frac{1}{2\ln2}\Upsilon,
\end{aligned}
\end{equation}
where
\begin{align}
&\Upsilon = \notag\\
&\exp\!\!\left(\!\! \frac{\rho\sigma_{\textrm{{rec}}}^2}{2(1\!-\!\rho)^2\sigma_{\textrm{{cov}}}^2(P|\tilde{h}|^2\!+\!\sigma_{\textrm{{A}}}^2)}\!\!\right)\!\!\operatorname{Ei}\!\!\left(\! \! \frac{\rho\sigma_{\textrm{{rec}}}^2}{2(1\!-\!\rho)^2\sigma_{\textrm{{cov}}}^2(P|\tilde{h}|^2\!+\!\sigma_{\textrm{{A}}}^2)}\!\!\right)\notag\\
&-\exp\!\left( \! \frac{(\rho\sigma_{\textrm{{A}}}^2\!+\!\sigma_{\textrm{{cov}}}^2)\sigma_{\textrm{{rec}}}^2}{2(1\!-\!\rho)^2P|\tilde{h}|^2\sigma_{\textrm{{A}}}^2\sigma_{\textrm{{cov}}}^2}\!\right)\operatorname{Ei}\!\left(\! \frac{(\rho\sigma_{\textrm{{A}}}^2\!+\!\sigma_{\textrm{{cov}}}^2)\sigma_{\textrm{{rec}}}^2}{2(1\!-\!\rho)^2P|\tilde{h}|^2\sigma_{\textrm{{A}}}^2\sigma_{\textrm{{cov}}}^2}\!\right),
\end{align}
where $\operatorname{Ei}(x)$ is the exponential integral.
Importantly, this approximation is asymptotically tight as $P\rightarrow\infty$.

\emph{Proof}: See Appendix A.
\label{proposition:prop1}
\end{prop}

\begin{figure*}[t]
\centering
\subfigure[$\sigma_{\textrm{{cov}}}^2=\sigma_{\textrm{{rec}}}^2=\sigma_{\textrm{{A}}}^2=1$.]{
\begin{minipage}[t]{0.32\linewidth}
\centering
\includegraphics[width=2.5in]{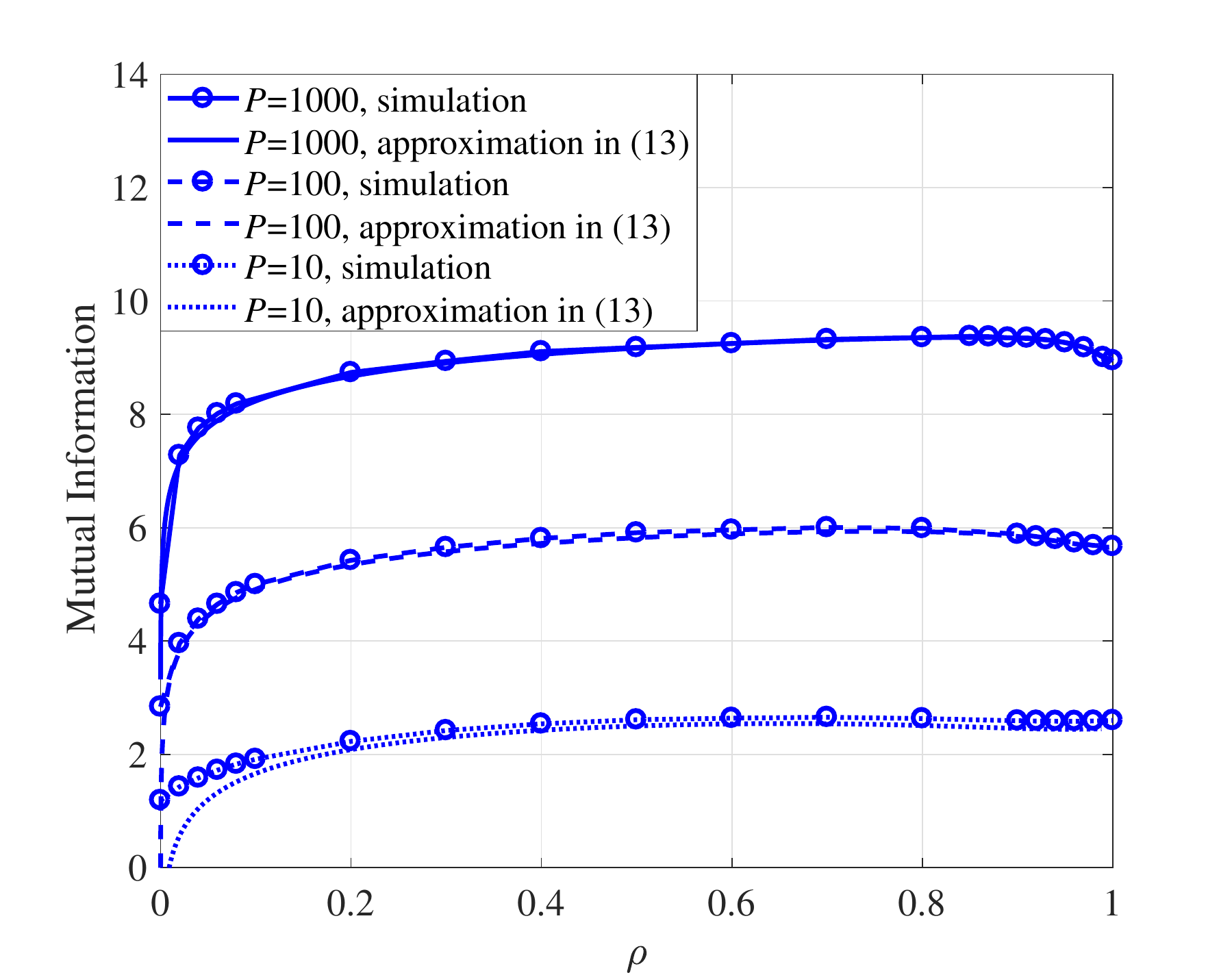}
\end{minipage}%
}%
\subfigure[$\sigma_{\textrm{{cov}}}^2=\sigma_{\textrm{{rec}}}^2=1,\sigma_{\textrm{{A}}}^2=0.1$.]{
\begin{minipage}[t]{0.32\linewidth}
\centering
\includegraphics[width=2.5in]{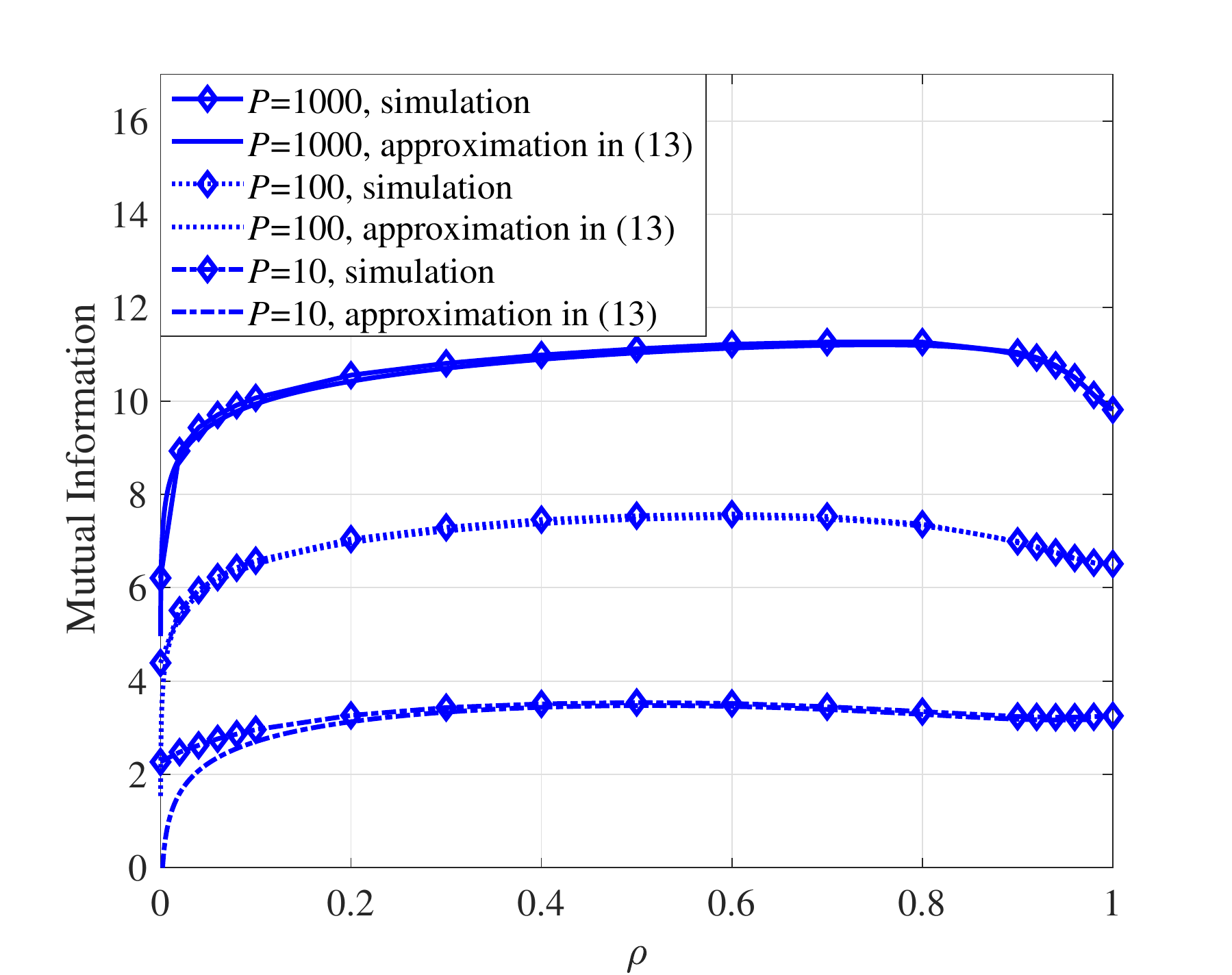}
\end{minipage}%
}%
\subfigure[$\sigma_{\textrm{{cov}}}^2=\sigma_{\textrm{{rec}}}^2=1$, $\sigma_{\textrm{{A}}}^2=0.01$.]{
\begin{minipage}[t]{0.32\linewidth}
\centering
\includegraphics[width=2.5in]{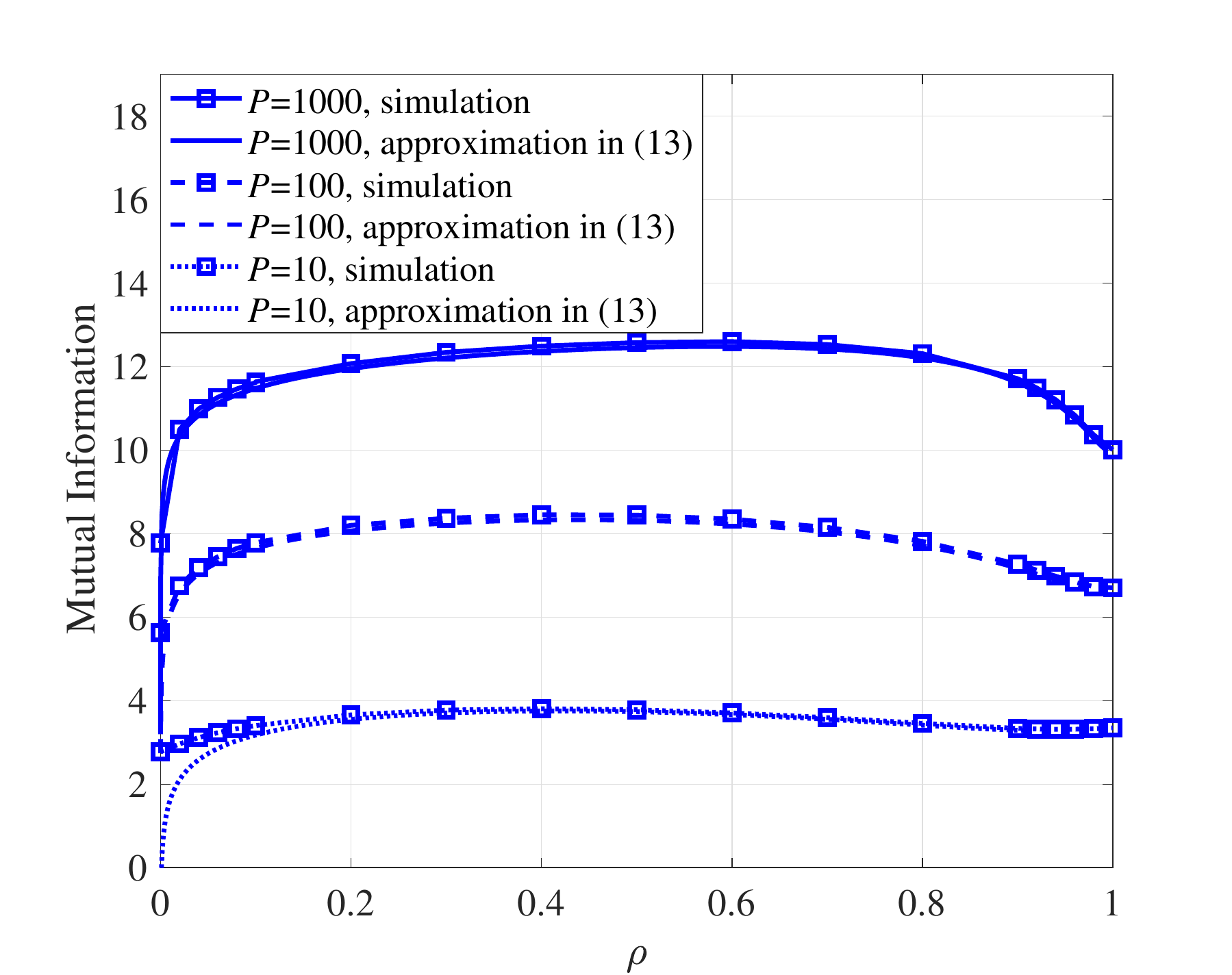}
\end{minipage}
}%
\centering
\caption{ Mutual information versus the power splitting ratio $\rho$.}
\label{fig:MI}
\end{figure*}

To verify the accuracy of the approximation in \eqnref{equ:MIhighSNR}, we plot the achievable mutual information against the power splitting ratio for different transmit power values in \figref{fig:MI}. Each subfigure differs in the antenna noise variance. The simulation result is obtained by using the Monte-Carlo-based histogram method for differential entropy estimation~\cite{michalowicz2013handbook}. Firstly, when comparing the simulation with the analytical approximation (in any subfigure), it is observed that the approximation is accurate even at moderate SNR, e.g., when $P = 10$, as long as the splitting ratio is not very small, e.g., $\rho>0.1$. As the SNR increases, the accuracy further improves even for small value of the splitting ratio. In short, Proposition~\ref{proposition:prop1} provides a very accurate approximation of the achievable mutual information at moderate to high SNR.

One can compare the values of mutual information at $\rho=0$ and $\rho=1$ to understand the performance difference between coherent detection and non-coherent PD-based detection. When the antenna noise is as large as the processing noises, i.e., \figref{fig:MI}~(a), the mutual information at $\rho=0$ is about half of that at $\rho=1$. On the other hand, when the antenna noise is much smaller than the processing noises, i.e., \figref{fig:MI}~(c), the mutual information values at $\rho=0$ and $\rho=1$ are much more closer. These observations are largely in agreement with prior studies on the performance of non-coherent PD-based receiver in~\cite[Equation (42)]{katz2004capacity} and~\cite[Proposition 8]{lapidoth2009capacity}, respectively.

In addition, one can also observe the trend in the optimal splitting ratio that maximizes the mutual information, that is, the optimal splitting ratio increases as SNR increases. For example, in \figref{fig:MI}~(b), the optimal $\rho$ is around 0.5 when $P=10$, and increases to 0.6 when $P=100$ and further increases to 0.8 when $P=1000$. Moreover, one can also observe the performance gain of the splitting receiver by comparing the mutual information achieved at the optimal $\rho$ with the mutual information achieved at either $\rho=0$ or $\rho=1$. Looking across all three subfigures, it is clear that the performance gain is very marginal when the antenna noise is as strong as the processing (i.e., conversion and rectifier) noises, while the performance gain is significant when the antenna noise is much weaker than the processing noises. To make such comparison more explicit, we will define metrics for measuring the performance gain in the next subsection.

\subsection{Mutual Information Performance Gain}
\label{section:MIgain}
One important question to answer is how much performance gain the splitting receiver brings in as compared to the traditional CD receiver or PD receiver. To answer this question, we first define two metrics:
\begin{defn}
The mutual information performance gain of the splitting receiver is
\begin{align}
\label{equ:MIgain}
&G_{\textrm{MI}}\triangleq \sup\big\{\mathcal{I}(\sqrt{P}|\tilde{h}|{{\tilde{X}}};{{{\tilde{Y}}}_1},{{{Y}}_2}): \rho\in (0,1)\big\}-\notag\\
&\max\big\{\mathcal{I}(\sqrt{P}|\tilde{h}|{{\tilde{X}}};{{{\tilde{Y}}}_1},{{{Y}}_2})|_{\rho=0},\mathcal{I}(\sqrt{P}|\tilde{h}|{{\tilde{X}}};{{{\tilde{Y}}}_1},{{{Y}}_2})|_{\rho=1}\big\},
\end{align}
where $\sup\{\cdot\}$ and $\max\{\cdot\}$ denote the supremum and maximum, respectively.

The value of $G_{\textrm{MI}}$ tells how much increase in mutual information that the splitting receiver can bring in as compared to the achievable mutual information by either the PD receiver or the CD receiver.
\label{define:MIde}
\end{defn}

\begin{defn}
The percentage gain of the splitting receiver is
\begin{align}
\label{equ:MIgainp}
&G_{\textrm{MI}}\%\triangleq \notag\\
&\frac{G_{\textrm{MI}}}{\max\!\big\{\!\mathcal{I}\!(\!\sqrt{P}|\tilde{h}|{{\tilde{X}}};\!{{{\tilde{Y}}}_1},\!{{{Y}}_2})|_{\rho=0},\mathcal{I}\!(\!\sqrt{P}|\tilde{h}|{{\tilde{X}}};\!{{{\tilde{Y}}}_1},\!{{{Y}}_2})|_{\rho=1}\!\big\}}\! \!\times \!\! 100\%.
\end{align}

Different from $G_{\textrm{MI}}$ which only tells the absolute performance gain, $G_{\textrm{MI}}\%$ tells the relative performance gain, which is more meaningful.
\label{define:MIde1}
\end{defn}

Before presenting our results, it is important to revisit the previous understanding on the mutual information performance of the splitting receiver as reported in~\cite{Liu2017A}. Specifically, the main result in~\cite{Liu2017A} on the mutual information performance gain in the asymptotically high SNR regime stated that the splitting receiver (with the optimal power splitting ratio) increases the mutual information by a factor of 1.5, in other words, $G_{\textrm{MI}}\% = 50\%$ as $P\rightarrow\infty$. This was an extremely encouraging result, but it was obtained under the key assumption of no antenna noise, i.e., $\sigma_{\textrm{{A}}}^2=0$. Consequently, one might think that the result of $G_{\textrm{MI}}\% = 50\%$ should be reasonably accurate when $\sigma_{\textrm{{A}}}^2$ is very small but not zero. Unfortunately, it turns out that the case of $\sigma_{\textrm{{A}}}^2=0$ and the case of $\sigma_{\textrm{{A}}}^2\neq0$ are fundamentally different, as shown in the following proposition.

\begin{prop}
In the asymptotically high SNR regime, the mutual information performance gain of the splitting receiver is given by
\begin{equation}
\label{equ:MIgain2}
\begin{aligned}
\mathop{\lim} \limits_{P\rightarrow\infty}G_{\textrm{MI}}= \frac{1}{2}\log_2\left(1+\frac{\sigma_{\textrm{{cov}}}^2}{\sigma_{\textrm{{A}}}^2}\right),
\end{aligned}
\end{equation}
\begin{equation}
\label{equ:MIgain3}
\begin{aligned}
\mathop{\lim} \limits_{P\rightarrow\infty}G_{\textrm{MI}}\%= 0\%,\,\,\,\,\text{for}\,\,\sigma_{\textrm{{A}}}^2\neq0.
\end{aligned}
\end{equation}
The optimal power splitting ratio, i.e., $\rho^*$, is arbitrarily close to 1 but not equal to 1 in the asymptotically high SNR regime.

\emph{Proof}: See Appendix B.\hfill
\label{proposition:prop2}
\end{prop}
\begin{table*}[t]
	\centering
	\caption{Optimal Power Splitting Ratio and Mutual Information Performance Gain}
	\begin{tabular}{|l|l|l|l|l|l|l|l|l|l|l|l|l|l|l|l|}
		\hline
		\multirow{2}{*}{Noise Conditions} & \multicolumn{3}{c|}{$P=10^2$} & \multicolumn{3}{c|}{$P=10^3$} & \multicolumn{3}{c|}{$P=10^4$} & \multicolumn{3}{c|}{$P=10^5$} & \multicolumn{3}{c|}{$P\rightarrow\infty$ (analytical)} \\ \cline{2-16}
		& \multicolumn{1}{c|}{$\rho^*$}             & \multicolumn{1}{c|}{$G_{\textrm{MI}}$}             & \multicolumn{1}{c|}{$G_{\textrm{MI}}\%$ }          & \multicolumn{1}{c|}{$\rho^*$ }              &\multicolumn{1}{c|}{$G_{\textrm{MI}}$}                &\multicolumn{1}{c|} {$G_{\textrm{MI}}\%$ }            & \multicolumn{1}{c|}{$\rho^*$ }              & \multicolumn{1}{c|}{$G_{\textrm{MI}}$}               & \multicolumn{1}{c|}{$G_{\textrm{MI}}\%$}             & \multicolumn{1}{c|}{$\rho^*$ }             & \multicolumn{1}{c|}{$G_{\textrm{MI}}$}               & \multicolumn{1}{c|}{$G_{\textrm{MI}}\%$}              &\multicolumn{1}{c|}{ $\rho^*$ }               &\multicolumn{1}{c|} {$G_{\textrm{MI}}$}              &\multicolumn{1}{c|} {$G_{\textrm{MI}}\%$ }            \\ \hline\hline
		\multicolumn{1}{|c|}{\begin{tabular}[c]{@{}c@{}}$\sigma_{\textrm{A}}^2=0.01$\\ $\sigma_{\textrm{cov}}^2=1$\\ $\sigma_{\textrm{rec}}^2=1$\end{tabular}} &  0.44             &  1.69            &   $25.4\%$           & 0.59              &    2.53           &  $25.4\% $           &  0.75             &   2.95            &  $22.2\% $           & 0.86              & 3.15              &  $19.0\%$            &   $\rightarrow 1$            &  3.33             &   \multicolumn{1}{c|}{0}             \\ \hline
		\multicolumn{1}{|c|}{\begin{tabular}[c]{@{}c@{}}$\sigma_{\textrm{A}}^2=0.01$\\ $\sigma_{\textrm{cov}}^2=1$\\ $\sigma_{\textrm{rec}}^2=0.1$\end{tabular}}       & 0.59              & 2.52              & $37.9\%$             &  0.75             & 2.95              &  $29.6\%$            & 0.86              &  3.15            &  $23.7\%$            &  0.93             &   3.24            &    $19.5\%$          &   $\rightarrow 1$            &   3.33            &     \multicolumn{1}{c|}{0}           \\ \hline
		\multicolumn{1}{|c|}{\begin{tabular}[c]{@{}c@{}}$\sigma_{\textrm{A}}^2=0.01$\\ $\sigma_{\textrm{cov}}^2=1$\\ $\sigma_{\textrm{rec}}^2=0.01$\end{tabular}}  & 0.75              & 2.93              &  $44.2\%$            & 0.86              &  3.14             & $31.5\%$             &  0.93             & 3.24              &  $24.4\%$            &0.96               &   3.29            &   $19.8\%$           &  $\rightarrow 1$             & 3.33              &  \multicolumn{1}{c|}{0}              \\ \hline\hline
		\multicolumn{1}{|c|}{\begin{tabular}[c]{@{}c@{}}$\sigma_{\textrm{A}}^2=0.01$\\ $\sigma_{\textrm{cov}}^2=1$\\ $\sigma_{\textrm{rec}}^2=1$\end{tabular}}  &  0.44            &  1.69             &   $25.4\%$           & 0.59              &    2.53           &  $25.4\% $           &  0.75            &   2.95           &  $22.2\% $           & 0.86             & 3.15             &  $19.0\%$            &   $\rightarrow 1$           &  3.33             &   \multicolumn{1}{c|}{0}                 \\ \hline
		\multicolumn{1}{|c|}{\begin{tabular}[c]{@{}c@{}}$\sigma_{\textrm{A}}^2=0.01$\\ $\sigma_{\textrm{cov}}^2=0.1$\\ $\sigma_{\textrm{rec}}^2=1$\end{tabular}}    & 0.52              & 0.27              & $2.73\%$             & 0.61              & 0.99              &  $7.54\%$            &  0.75             & 1.38              &  $8.35\%$            &  0.86             &   1.56            &    $7.88\%$          & $\rightarrow 1$              &   1.73            &   \multicolumn{1}{c|}{0}             \\ \hline
		\multicolumn{1}{|c|}{\begin{tabular}[c]{@{}c@{}}$\sigma_{\textrm{A}}^2=0.01$\\ $\sigma_{\textrm{cov}}^2=0.01$\\ $\sigma_{\textrm{rec}}^2=1$\end{tabular}}   &  0.60            & 0.02              & $0.24\%$             &  0.69             &   0.04            &  $0.26\%$            & 0.75              &   0.27           & $1.43\%$             & 0.85             &  0.39            &  $1.75\%$            &    $\rightarrow 1$           &  0.50             & \multicolumn{1}{c|}{0}               \\ \hline
	\end{tabular}
	\label{tab:MIan}
\end{table*}

\begin{remark}
The splitting receiver with the antenna noise in \figref{fig:Fig1sysmod} can be treated as a cascade of an AWGN channel (where the signal is affected by the antenna noise before entering the splitter) and an antenna-noise-free splitting channel (where the signal is split into the two streams each affected by the corresponding processing noise). The splitting channel was thoroughly analyzed in~\cite{Liu2017A}, where the achievable mutual information was shown to have a scaling law of $\frac{3}{2}\log_2(P)$, hence the 50\% gain reported therein. When the antenna noise is considered, the achievable mutual information is upper bounded by the minimum of the two cascaded channels~\cite{BookInfo}. It is well known that the AWGN channel has a scaling law of $\log_2(P)$, hence the mutual information scaling law of the overall channel is limited to $\log_2(P)$. Therefore, the percentage gain in mutual information approaches zero asymptotically, as stated in Proposition~\ref{proposition:prop2}. This is why the existence of the antenna noise fundamentally changes the behavior of achievable mutual information of the splitting receiver. Nevertheless, the scaling law result does not reveal any possible absolute improvement in the mutual information. We know from~\cite{Liu2017A} that the splitting channel increases the achievable mutual information, hence, intuitively there can still be an absolute gain. The result in Proposition~\ref{proposition:prop2} has confirmed that there is an absolute gain in mutual information which approaches to a constant value as SNR approaches infinity. In addition, the optimal power splitting ratio approaches one as SNR increases, which is again very different from the previous result in~\cite{Liu2017A} which stated that $\rho^{*}\rightarrow 1/3$ as SNR approaches infinity. Therefore, the previous understanding in~\cite{Liu2017A} is only valid for the special and impractical case of no antenna noise and it cannot be used to infer the mutual information performance and the optimal design of the splitting receiver in any practical case where the antenna noise always exists.
\end{remark}

\begin{remark}
Note that it may be counterintuitive to see that $\rho^*$ gets close to 1 while there is still an asymptotically constant gap between the maximum mutual information at $\rho^*$ and the mutual information achieved at $\rho=1$. This is actually due to the behavior of the mutual information as a function of $\rho$ at high SNR. More specifically, as $\rho$ increases, the mutual information increases to its maximum value at $\rho^*$ (which is close to 1), and then drops sharply as $\rho$ further increases till reaching 1. This behavior has been numerically observed (results not shown for brevity). Hence, the mutual information is left continuous at $\rho = 1$ and the sharp drop of mutual information around this small region of $\rho$ causes the asymptotically constant gap reported in the proposition.
\end{remark}

The result in Proposition~\ref{proposition:prop2} appears discouraging, because it implies that asymptotically there is no performance improvement by using the splitting receiver. However, the numerical results in Fig.~\ref{fig:MI} do show some notable performance improvement at practical SNR values. Therefore, it is important to investigate the performance improvement at practical SNR values. To this end, Table~\ref{tab:MIan} shows the results on the optimal power splitting ratio $\rho^*$, mutual information gain $G_{\textrm{MI}}$ and its percentage gain $G_{\textrm{MI}}\%$ for different values of the transmit power $P$. It considers a baseline case of $\sigma_{\textrm{{rec}}}^2=\sigma_{\textrm{{cov}}}^2=1$, $\sigma_{\textrm{{A}}}^2=0.01$ (repeated in the first and fourth rows of Table~\ref{tab:MIan}), where the processing noises are much stronger than the antenna noise, because this is the scenario where the splitting receiver gives significant performance gain. Then, different rectifier noise variances $\sigma_{\textrm{{rec}}}^2=1,\ 0.1$ and $0.01$ are used in the first, second and third rows of Table~\ref{tab:MIan}, and different conversion noise variances $\sigma_{\textrm{{cov}}}^2=1,\ 0.1$ and $0.01$ are used in the fourth, fifth and sixth rows of Table~\ref{tab:MIan}.

We first discuss the trend in the optimal splitting ratio $\rho^*$. Looking horizontally across each row in Table~\ref{tab:MIan}, there is a clear trend of increasing $\rho^*$ as SNR increases. Nevertheless, the value of $\rho^*$ at finite SNR does not reach the neighborhood of its asymptotic value of 1, even when $P=10^5$. Next we discuss the mutual information gain. Looking across each row, we again see that $G_{\textrm{MI}}$ increases towards its asymptotic value as SNR increases. In terms of the percentage gain, $G_{\textrm{MI}}\%$ may or may not initially increase with SNR but eventually reduces as SNR increases further. More importantly, we observe some huge percentage gains at moderate SNR values, e.g., $G_{\textrm{MI}}\%$ can go as high as $44.2\%$ when $P=10^2$ and the percentage gains still stay significant even at very high SNR values, e.g., $G_{\textrm{MI}}\%$ can be as high as $19.8\%$ when $P=10^5$. Hence, the pessimistic asymptotic result of $G_{\textrm{MI}}\%=0\%$ is not indicative for practical scenarios. In fact, the splitting receiver can provide very significant performance improvement at practical SNRs.

Another implication of Proposition~\ref{proposition:prop2} is that the mutual information performance gain is primarily determined by the relative strength of the conversion noise versus the antenna noise, while the strength of the rectifier noise has negligible effect. Comparing $G_{\textrm{MI}}\%$ among the first, second and third rows in Table~\ref{tab:MIan}, where $\sigma_{\textrm{{rec}}}^2$ (i.e., the strength of the rectifier noise) is changed, we see that the performance gain stays roughly the same for different values of $\sigma_{\textrm{{rec}}}^2$ when the SNR is sufficiently high, e.g., $P=10^5$, which agrees with the analytical result in Proposition~\ref{proposition:prop2}.  Comparing $G_{\textrm{MI}}\%$ among the fourth, fifth and sixth rows in Table~\ref{tab:MIan}, where $\sigma_{\textrm{{cov}}}^2$ is changed (hence the relative strength of the conversion noise versus the antenna noise is changed), we see that the performance gain differs significantly even at very high SNRs, which also agrees with the insight obtained from Proposition~\ref{proposition:prop2}. Overall, our analytical and numerical results give us important insights into the scenarios in which the splitting receiver has notable advantages, as compared to the traditional receivers.

It should also be noted that the results on the achieved mutual information of the splitting receiver are obtained under the assumption of Gaussian input distribution. We know from the existing literature that the optimal input distribution for the traditional CD receiver is Gaussian~\cite{BookInfo}, while the optimal input distribution for the PD receiver is still unknown but certainly non-Gaussian~\cite{katz2004capacity,lapidoth2009capacity}. Therefore, finding the optimal input distribution for the splitting receiver is a very challenging open problem. Although we do not show the numerical results here for brevity, we have numerically investigated other input distributions and found that, for example, some Gaussian mixture distributions can achieve higher mutual information than the standard Gaussian distribution at some moderate SNR values we tested. Therefore, the mutual information performance of the splitting receiver can generally be even better than what has been illustrated if a better input distribution is used.

\section{Signal Detection Rule and Performance}
In this section, we switch our attention from information theoretic performance with Gaussian input to signal detection with practical modulation schemes. We assume that the transmitted symbols are drawn from a two-dimensional constellation where each symbol/point in the constellation has equal probability to be drawn. Denote the set of all constellation points as $\Omega_{\textrm{gen}}$. In what follows, we will first present the optimal detection rule, i.e., the ML detection. Then, we will present a low-complexity detection rule for practical purposes. Finally, we will present results in terms of SER.

\subsection{Optimal Detection Rule}
\label{section:secoptdection}
Since all constellation points are of equal probability, the ML detector is the optimal signal detection method. Specifically, the ML detection output is given by
\begin{equation}
\label{equ:MLopt}
\hat{X}=\mathop{\arg\max} \limits_{\tilde{X}\in \Omega_{\textrm{gen}}} f_{\tilde{Y}_{1}, Y_2}\big(\tilde{y}_{1}, y_2|\tilde{X}\big),
\end{equation}
where $f_{\tilde{Y}_{1}, Y_2}\big(\tilde{y}_{1}, y_2|\tilde{X}\big)$ is the conditional PDF of the received signal/symbol given that the transmitted symbol is $\tilde{X}$. Here, we call $(\tilde{Y}_{1}, Y_2)$ the received symbol. However, one should note that the actual received symbol as shown in \figref{fig:Fig1sysmod} is $(\tilde{Y}^{'}_{1}, Y^{'}_2)$.
For convenience of subsequent analysis, we explicitly write out the real part and imaginary part of $\tilde{Y}_1$, denoted as $Y_{1r}$ and $Y_{1i}$. Based on \eqnref{equ:CDreceiver} and \eqnref{equ:PDreceiverch}, this conditional PDF can be written as
\begin{align}
\label{equ:3Dpdf}
&f_{{Y}_{1r},{Y}_{1i}, Y_2}\big({y}_{1r}, {y}_{1i}, y_2|\tilde{X}\big)\notag\\ &=\int_{{W}_r}\int_{{W}_i}f_{{Y}_{1r},{Y}_{1i},Y_2}\big({y}_{1r},{y}_{1i},y_2|{w}_r,{w}_i,\tilde{X}\big)\notag\\
&\,\,\,\,\,\, f_{{\tilde{W}}}\big({w}_r,{w}_i\big)\,\mathrm{d}{w}_i\mathrm{d}{w}_r,
\end{align}
where we have also explicitly written out the real part and imaginary part of the antenna noise $\tilde{W}$ denoted as ${W}_r$ and ${W}_i$, respectively. Hence, $f_{{{\tilde{W}}}}({w}_r,{w}_i)$ denotes the PDF of the antenna noise, which is given as
\begin{equation}
\label{equ:noiseW}
f_{{\tilde{W}}}\big({{{w}}}_r,{{{w}}}_i\big)
=\frac{1}{\pi\sigma_{\textrm{A}}^2}\exp\left(-\frac{
{{{w}}}_r^2+{{{w}}}_i^2}{\sigma_{\textrm{A}}^2}\right).
\end{equation}

Note that
$f_{{Y}_{1r},{Y}_{1i},Y_2}\big({y}_{1r},{y}_{1i},y_2|{w}_r,{w}_i,\tilde{X}\big)$
is the conditional PDF of the received symbol when the antenna noise and transmitted symbol are given, which is derived as \eqref{equ:conpdf1}.
\newcounter{mytempeqncnt}
\begin{figure*}[!b]
\normalsize
\setcounter{MYtempeqncnt}{\value{equation}}
\setcounter{equation}{21}
\hrulefill
\begin{equation}
\label{equ:conpdf1}
\begin{aligned}
f_{{Y}_{1r},{Y}_{1i},Y_2}\big({y}_{1r},{y}_{1i},y_2|{w}_r,{w}_i,\tilde{X}\big)&=
\frac{1}{\pi\sigma_{\textrm{cov}}^2}\exp\left(-\frac{\left({{{{y}}}_{1r}}-\sqrt{\rho}\left(\sqrt{P}|\tilde{h}|{{{X}}}_{r}+{{{w}}}_r\right)\right)^2}{\sigma_{\textrm{cov}}^2}-
\frac{\left({{{{y}}}_{1i}}
-\sqrt{\rho}\left(\sqrt{P}|\tilde{h}|{{{X}}}_i+{{{w}}}_i\right)\right)^2}{\sigma_{\textrm{cov}}^2}\right)\\
&\frac{1}{\sqrt{2\pi\sigma_{\textrm{rec}}^2}}\exp\left(-\frac{\left(
{{{y}}_2}
-\left(\left(1-\rho\right)\left(\left(\sqrt{P}|\tilde{h}|{{{X}}}_r+{{{w}}}_r\right)^2+\left(\sqrt{P}|\tilde{h}|{{{X}}}_i+{{{w}}}_i\right)^2\right)\right)\right)^2}{2\sigma_{\textrm{rec}}^2}\right).
\end{aligned}
\end{equation}
\setcounter{equation}{\value{MYtempeqncnt}}
\vspace*{4pt}
\end{figure*}
\setcounter{equation}{22}

Using \eqnref{equ:noiseW} and \eqnref{equ:conpdf1}, the receiver is able to evaluate the conditional PDF in \eqref{equ:3Dpdf} for each candidate transmit symbol $\tilde{X}$, and hence, determine the detection output that maximizes the conditional PDF.
\begin{remark}
The complexity of the optimal ML detector is high due to the double integrals in the likelihood function in \eqref{equ:3Dpdf} which has no closed-form expression. The receiver needs to either evaluate this likelihood function for all received symbols or construct the decision regions for all candidate transmit symbols based on the likelihood function. This has to be done for a given operating condition, e.g., the current operating SNR, and redone as soon as the operating condition changes. Hence, the requirement of numerical evaluation of double integrals significantly increases the complexity of the detection. It is, therefore, important to derive a low-complexity detection rule for practical purposes.
\end{remark}

\subsection{Low-Complexity Detection Rule}
\label{section:seclowdection}
In order to derive a low-complexity detection rule, we first apply a linear scaling on $Y_2$ in \eqnref{equ:PDreceiverch}, i.e., dividing $Y_2$ by $\sqrt{P}$. Note that applying such scaling does not change the signal quality nor the performance of the receiver, and there is no loss of generality. With a slight abuse of notation, we keep the same notation for the scaled signal as
\begin{align}
\label{equ:PDreceiverchC}
{{{Y}}_2} =
(1-\rho)|\tilde{h}|^2\sqrt{P}\biggl(|{{\tilde{X}}}|^2+&\frac{2{X}_r{{{W}}}_r}{\sqrt{P}|\tilde{h}|}+\frac{2{X}_i{{{W}}}_i}{\sqrt{P}|\tilde{h}|}\notag\\
&+\frac{|{{\tilde{W}}}|^2}{{{P}|\tilde{h}|^2}}\biggr)+\frac{{{N}}}{\sqrt{P}}.
\end{align}

As $P$ increases, the higher-order noise term $\frac{|{{\tilde{W}}}|^2}{{{P}|\tilde{h}|^2}}$ becomes negligible as compared to the other first-order noise terms $\frac{2{X}_r{{{W}}}_r}{\sqrt{P}|\tilde{h}|}$ and $\frac{2{X}_i{{{W}}}_i}{\sqrt{P}|\tilde{h}|}$. Hence asymptotically we can approximate $Y_2$ in \eqref{equ:PDreceiverchC} by dropping the higher-order noise term, that is
\begin{align}
\label{equ:PDreceiverch2C1}
\mathop{\lim} \limits_{P\rightarrow\infty}{{{Y}}_2} = (1-\rho)\sqrt{P}&\biggl(|\tilde{h}|^2|{{\tilde{X}}}|^2+\frac{2|\tilde{h}|{X}_r{{{W}}}_r}{\sqrt{P}}\notag\\
&+\frac{2|\tilde{h}|{X}_i{{{W}}}_i}{\sqrt{P}}\biggr)+\frac{{N}}{\sqrt{P}}.
\end{align}

For convenience, we define ${{N}}_s\triangleq\frac{{N}}{\sqrt{P}}$, which follows $\mathcal{N} (0, \sigma_{{N}_s}^2)$, where $\sigma_{{N}_s}^2=\frac{\sigma_{\textrm{rec}}^2}{P}$.

Based on the expressions of $\tilde{Y}_1$ in \eqnref{equ:CDreceiver} and the approximated $Y_2$ in \eqnref{equ:PDreceiverch2C1}, we obtain the following low-complexity detection rule.

\begin{prop}
A low-complexity detection rule is given by
\begin{equation}
\label{equ:MLsub2opt2}
\hat{X}=\mathop{\arg\max} \limits_{\tilde{X}\in \Omega_{\textrm{gen}}} \hat{f}_{\tilde{Y}_{1},
Y_2}\big(\tilde{y}_1,y_2|\tilde{X}\big),
\end{equation}
where
\begin{align}
\label{equ:MLsub2opt1}
&\hat{f}_{\tilde{Y}_{1}, Y_2}\big(\tilde{y}_1,y_2|\tilde{X}\big)=\notag\\
&\frac{1}{\sqrt{2\pi^{3}(\rho{\sigma_{\textrm{A}}^2}\!+\!{{\sigma_{\textrm{cov}}^2}})\big({{\sigma_{\textrm{cov}}^2}}{\sigma_{{N}_s}^2}\!+\!2{\sigma_{\textrm{A}}^2}{{\sigma_{\textrm{cov}}^2}}|\tilde{X}|^2(\rho\!-\!\!1)^2\!+\!\rho{\sigma_{\textrm{A}}^2}{\sigma_{{N}_s}^2}\big)}}\notag\\
&\exp\!\!\left(\!\!-\frac{\big(2\sqrt{\rho}{\sigma_{\textrm{A}}^2}(\rho\!-\!\!1)({{{X}}_r}T_{1r}\!+\!{{{X}}_i}T_{1i})\!+\!(\rho{\sigma_{\textrm{A}}^2}\!+\!{{\sigma_{\textrm{cov}}^2}})T_2\big)^2}
{2(\rho{\sigma_{\textrm{A}}^2}\!+\!{{\sigma_{\textrm{cov}}^2}})\!\big(\!{{\sigma_{\textrm{cov}}^2}}{\sigma_{{N}_s}^2}\!\!+\!2{\sigma_{\textrm{A}}^2}{{\sigma_{\textrm{cov}}^2}}|\tilde{X}|^2(\rho\!-\!\!1)^2\!\!+\!\!\rho{\sigma_{\textrm{A}}^2}{\sigma_{{N}_s}^2}\!\big)}\!\!\right)\notag\\
&\exp\left(-\frac{|T_{1r}|^2+|T_{1i}|^2}{\rho{\sigma_{\textrm{A}}^2}+{{\sigma_{\textrm{cov}}^2}}}\right)
\end{align}
is the conditional joint PDF of $\tilde{Y}_1$ in \eqref{equ:CDreceiver} and approximated $Y_2$ in \eqref{equ:PDreceiverch2C1},
and $T_{1r}={{{{y}}}_{1r}}
-\sqrt{\rho}\sqrt{P}|\tilde{h}|{{{X}}_r}$, $T_{1i}={{{{y}}}_{1i}} -\sqrt{\rho}\sqrt{P}|\tilde{h}|{{{X}}_i}$, $T_2={{{y}}_2}
- (1-\rho)\sqrt{P}|\tilde{h}|^2|{{\tilde{X}}}|^2$. Importantly, this detection rule is asymptotically optimal as $P\rightarrow\infty$.

\emph{Proof}: See Appendix C.
\label{proposition:prop3}
\end{prop}


\begin{remark}
As compared with the optimal ML detection rule where the likelihood function has a double integral, the low-complexity detection rule has a closed-form likelihood function which can be evaluated quickly. This significantly reduces the computational complexity of signal detection at the receiver. The low-complexity detection rule is asymptotically optimal at high SNRs because it is obtained by ignored an insignificant noise term while keeping all the significant noise terms in the signal model.
\end{remark}

\subsection{SER Performance}

In this subsection, we study the SER performance of the splitting receiver. Despite having a low-complexity detection rule derived in \secref{section:seclowdection}, it is generally difficult to obtain analytically tractable characterization for SER performance. In trivial cases such as phase shift keying (PSK) where all symbols in the constellation have the same magnitude, it is clear that the optimal splitting ratio is 1, and hence, the SER performance is given by well-known analytical results for the traditional CD receiver. However, for non-trivial cases such as QAM, the optimal splitting ratio is not necessarily 1 and finding a mathematically tractable and useful SER expression is extremely difficult even at high SNRs. Therefore, we resort to numerical investigation to study the SER performance and present the results for rectangular 64-QAM.

\begin{figure*}[t]
\centering
\subfigure[$\sigma_{\textrm{{cov}}}^2=\sigma_{\textrm{{rec}}}^2=\sigma_{\textrm{{A}}}^2=1$.]{
\begin{minipage}[t]{0.32\linewidth}
\centering
\includegraphics[width=2.515in]{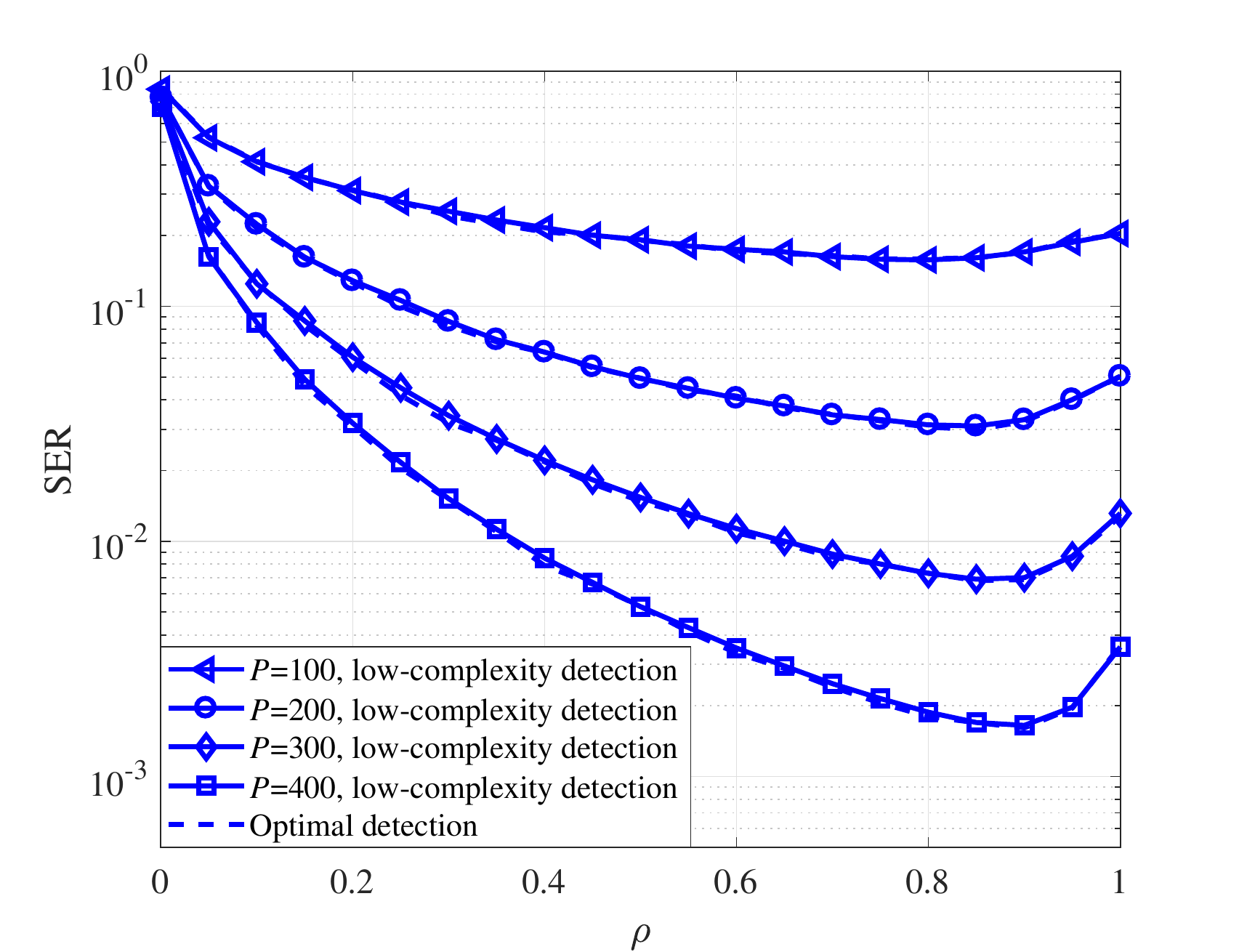}
\end{minipage}%
}%
\subfigure[$\sigma_{\textrm{{cov}}}^2=\sigma_{\textrm{{rec}}}^2=1,\sigma_{\textrm{{A}}}^2=0.1$.]{
\begin{minipage}[t]{0.32\linewidth}
\centering
\includegraphics[width=2.472in]{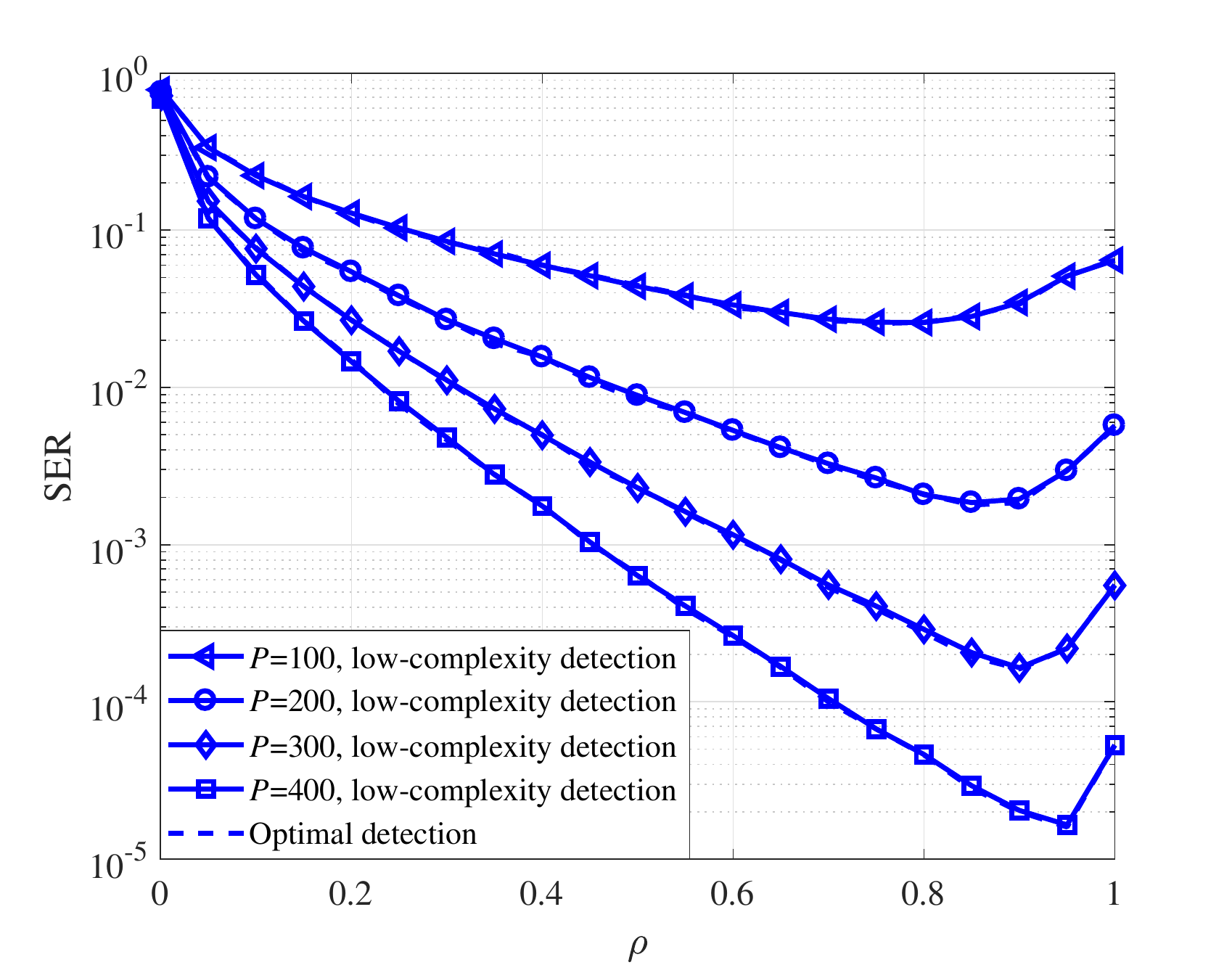}
\end{minipage}%
}%
\subfigure[$\sigma_{\textrm{{cov}}}^2=\sigma_{\textrm{{rec}}}^2=1$.]{
\begin{minipage}[t]{0.32\linewidth}
\centering
\includegraphics[width=2.56in]{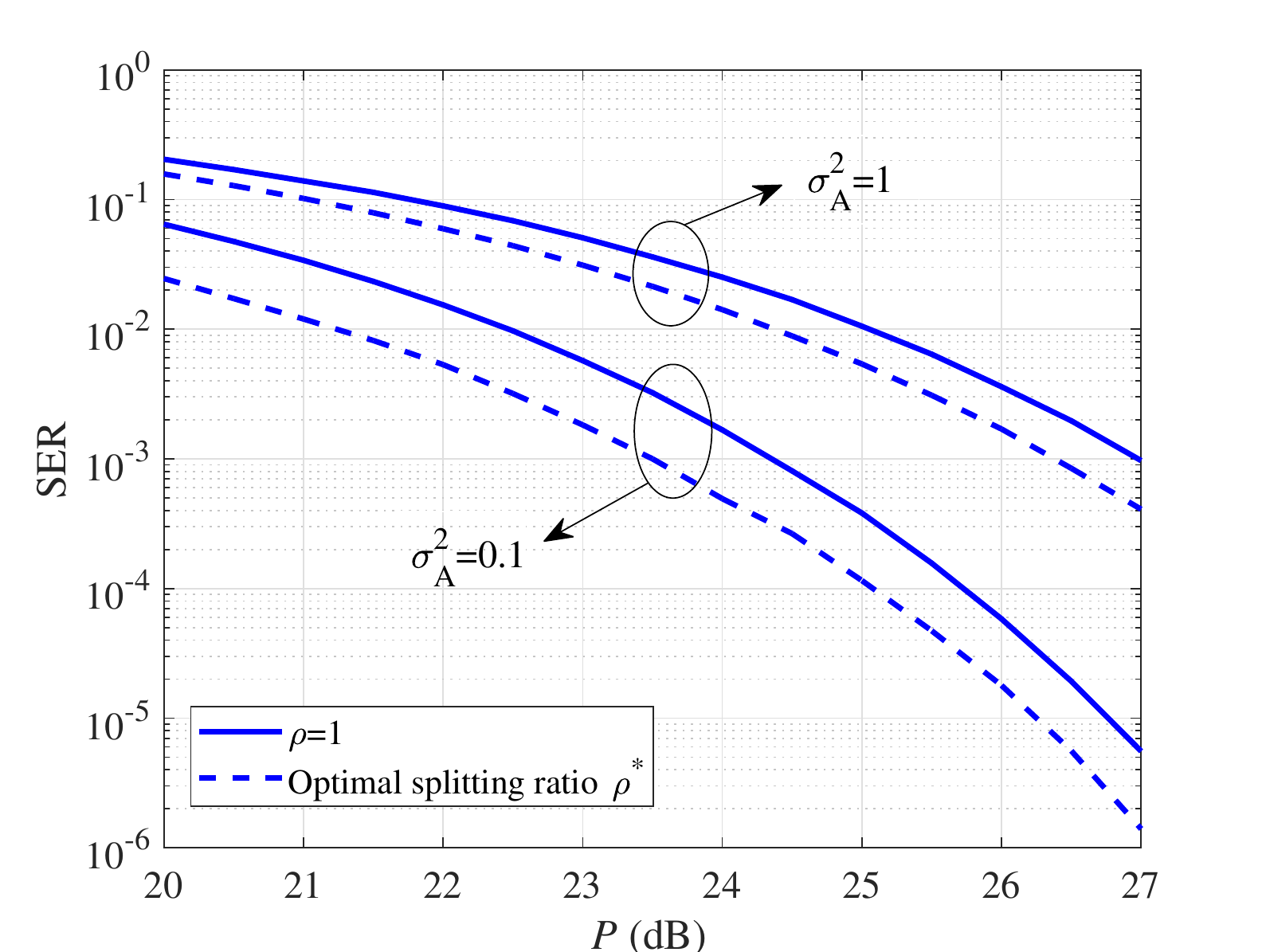}
\end{minipage}
}
\centering
\caption{SER versus the power splitting ratio $\rho$ or the transmit power $P$ for 64-QAM.}
\label{fig:SERQAM}
\end{figure*}

\figref{fig:SERQAM}~(a) and \figref{fig:SERQAM}~(b) show the SER performance versus the power splitting ratio $\rho$ for two different noise conditions. In each subfigure, we plot both the SER achieved by the low-complexity detection rule (in \secref{section:seclowdection}) indicated using solid lines, as well as the SER achieved by the optimal detection rule (in \secref{section:secoptdection}) indicated using dashed lines. We see almost no performance difference between the low-complexity detection and the optimal detection for the illustrated range of SERs. We have also done further numerical investigations (results not shown here for brevity) over a wider range of $P$ values and again confirmed the indistinguishable performance between the two detection rules. This also agrees with the analytical result in Proposition~\ref{proposition:prop3} that the low-complexity detection rule is asymptotically optimal at high SNR.

Focusing on the optimal splitting ratio (denoted by $\rho^*$) that minimizes SER, we see that $\rho^*$ increases as $P$ increases. For example, in \figref{fig:SERQAM}~(a), the value of $\rho^*$ increases from 0.8 to 0.9 when $P$ increases from 100 to 400. Similar trend is also observed in  \figref{fig:SERQAM}~(b).
For large values of $P$, the SER curve reaches its minimum at $\rho^*$ whose value is close to 1, and then the SER curve goes up (sharply) after $\rho$ passes its optimal value.
From these observations, we conjecture that, as $P$ increases, the value of $\rho^*$ that minimizes SER will approach 1 but never reaches 1. This is consistent with the analytical results we obtained on $\rho^*$ that maximizes the mutual information (in \secref{section:MIgain}).

To illustrate the advantage of the splitting receiver, we compare the minimum SER achieved by the splitting receiver (i.e., with $\rho^*$) and the SER achieved by the traditional CD receiver (i.e., $\rho = 1$). One way to make such a comparison is to directly look at the SER values at a given value of $P$. Let us look at the SER curves for $P=200$ as an example. In  \figref{fig:SERQAM}~(a), the splitting receiver achieves SER~$\approx 3\times10^{-2}$ and the CD receiver achieves SER~$ \approx 5\times10^{-2}$, i.e., 1.67 times more errors. In  \figref{fig:SERQAM}~(b), the splitting receiver achieves SER~$\approx 2\times10^{-3}$ and the CD receiver achieves SER~$\approx 6\times10^{-3}$, i.e., 3 times more errors. Clearly, the advantage of the splitting receiver is enhanced when the antenna noise is weaker. Another way of comparison is to look at the transmit power consumption for achieving the same SER. To this end, we re-plot the SER versus $P$ (in dB scale for showing the ratio of power consumption) in  \figref{fig:SERQAM}~(c). Again, we see that the advantage of the splitting receiver is enhanced when the antenna noise is weaker. Focusing on the two curves for $\sigma_{\textrm{A}}^2=0.1$, the difference in $P$ for achieving a target SER of $10^{-2}$ is 1.2 dB (i.e., the CD receiver requires 32\% more transmit power). This difference in $P$ is 0.7 dB (i.e., the CD receiver requires 17\% more transmit power) if the target SER is $10^{-4}$. It is evident that the use of the splitting receiver provides a notable performance gain for practically relevant range of SER values.

\begin{remark}
Apart from 64-QAM, we have numerically investigated other modulation orders and other types of modulations (results not shown in the paper). We have observed that the performance advantage of the splitting receiver, as compared to the traditional CD receiver, behaves very differently for different types of modulations. Even for the same type of modulation, e.g., amplitude and phase shift keying (APSK), different constellation designs can lead to very different results. For example, we have considered two 32-APSK constellation designs that are specified for the DVB-S2X standard~\cite{etsi2014302}: one has three rings with 4, 12, and 16 points on each ring, and the other has four rings with 4, 8, 4, and 16 points on each ring. Our numerical investigation shows that the CD receiver is optimal (i.e., $\rho^* = 1$) for the (4, 12, 16) design of 32-APSK, while the splitting receiver achieves a significant performance gain for the (4, 8, 4, 16) design of 32-APSK and, interestingly, the gain is more profound at larger $P$ or lower SER. Therefore, the performance analysis and characterization of the splitting receiver for practical modulation can only be done in a case-by-case basis. This also leads to an important open problem: how to design the best modulation or constellation diagram for the splitting receiver?
\label{remark:Re4}
\end{remark}

\section{Conclusions and Future Work}

In this paper, we have re-examined the splitting receiver proposed in~\cite{Liu2017A} and studied its performance by considering the antenna noise that was ignored in~\cite{Liu2017A}. We have shown that the inclusion of the antenna noise fundamentally changes the understanding on the information-theoretic performance of the splitting receiver. For signal detection with practical modulations, we have derived a low-complexity detection rule that is shown to achieve near-optimal detection performance. Our numerical results on both the mutual information performance and the SER performance have demonstrated the advantage of using the splitting receiver (as compared to the traditional CD receiver) and provided insights on the conditions under which the performance advantage is most profound. As mentioned in Remark~\ref{remark:Re4}, one important future work is to design suitable modulations (i.e., constellation diagrams) for the splitting receiver, in order to fully exploit the benefit of this new receiver architecture.

\begin{appendices}

\section{ Proof of Proposition 1 }
Based on the property of mutual information invariance under scaling of random variables \cite{BookInfo}, we apply further linear scaling on the received signals in \eqnref{equ:CDreceiver} and \eqnref{equ:PDreceiverch}. With some slight abuse of notation, the further scaled received signals are given by
\setcounter{equation}{0}
\renewcommand\theequation{A.\arabic{equation}}
\begin{align}
\label{equ:spre}
&{{{\tilde{Y}}}_1} = \sqrt{\Theta_1}\left({{\tilde{X}}}+\frac{{{\tilde{W}}}}{\sqrt{P}|\tilde{h}|}\right)+\frac{{{\tilde{Z}}}}{\sqrt{P}|\tilde{h}|},\notag\\
&{{{Y}}_2} = \sqrt{\Theta_2}k\sqrt{P}|\tilde{h}|\biggl|{{\tilde{X}}}+\frac{{{\tilde{W}}}}{\sqrt{P}|\tilde{h}|}\biggr|^2+k\frac{{{N}}}{\sqrt{P}|\tilde{h}|},
\end{align}
where $k\triangleq\frac{\sigma_{\textrm{cov}}}{\sqrt{2}\sigma_{\textrm{rec}}}$, $\Theta_1\triangleq \rho $, and $\Theta_2\triangleq (1-\rho)^2 $.
Thus, it is easy to verify that the real and imaginary parts of $\frac{{{\tilde{Z}}}}{\sqrt{P}|\tilde{h}|}$ and $k\frac{{{N}}}{\sqrt{P}|\tilde{h}|}$ are independent with each other and follow the same distribution $\mathcal{N}\big(0,\frac{\sigma_{\textrm{cov}}^2}{2P|\tilde{h}|^2}\big)$.

In the following, with some slight abuse of notation, $\sqrt{P}|\tilde{h}|$ and $P|\tilde{h}|^2$ are denoted as $\sqrt{P}$ and $P$, respectively, for simplicity.

We define the following random variables as
\begin{equation}
\label{equ:Variables1}
{{{\tilde{X}}}_1} = \sqrt{\Theta_1}{{\tilde{X}}},
\end{equation}
\begin{equation}
\label{equ:Variables2} {{{X}}_2} = \sqrt{\Theta_2}k\sqrt{P}|{{\tilde{X}}}|^2,
\end{equation}
\begin{equation}
\label{equ:Variables3}
{{{\tilde{X}}}'_1} = \sqrt{\Theta_1}\left({{\tilde{X}}}+\frac{{{\tilde{W}}}}{\sqrt{P}}\right),
\end{equation}
\begin{equation}
\label{equ:Variables4}
{{{X}}'_2} = \sqrt{\Theta_2}k\sqrt{P}\biggl|{{\tilde{X}}}+\frac{{{\tilde{W}}}}{\sqrt{P}}\biggr|^2.
\end{equation}

Because of the Markov chain $\sqrt{P}{{\tilde{X}}}\rightarrow ({{{\tilde{X}}}_1}, {{{X}}_2})\rightarrow ({{{\tilde{X}}}'_1}, {{{X}}'_2})\rightarrow ({{{\tilde{Y}}}_1}, {{{Y}}_2})$ and the smooth and uniquely invertible map from $\sqrt{P}{{\tilde{X}}}$ to  $({{{\tilde{X}}}_1}, {{{X}}_2})$, we have
\begin{equation}
\label{equ:MI}
\mathcal{I}(\sqrt{P}{{\tilde{X}}};{{{\tilde{Y}}}_1}, {{{Y}}_2})=\mathcal{I}({{{\tilde{X}}}_1}, {{{X}}_2};{{{\tilde{Y}}}_1}, {{{Y}}_2}).
\end{equation}

\begin{figure}[t]
	\centering
	\includegraphics[scale=0.8]{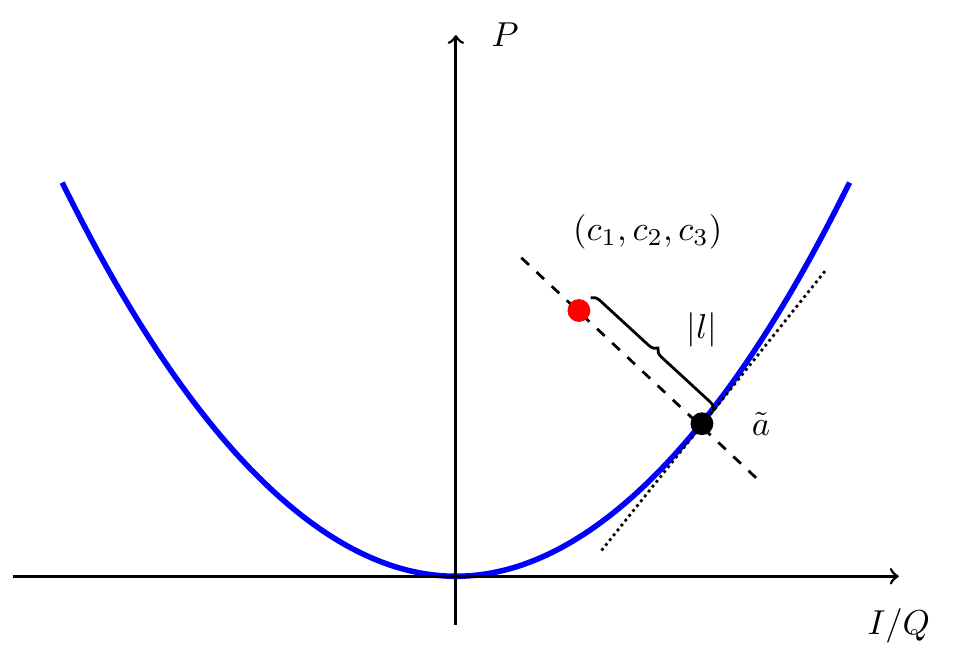}	
	\caption{Illustration of the PN coordinate system (two-dimensional illustration for simplicity).}
	\label{fig:illustration}	
\end{figure}

To analyze $\mathcal{I}({{{\tilde{X}}}_1}, {{{X}}_2};{{{\tilde{Y}}}_1}, {{{Y}}_2})$, we use paraboloid-normal (PN) coordinate system proposed in~\cite{Liu2017A} as illustrated in Fig.~\ref{fig:illustration}, which is based on a paraboloid $\mathcal{U}$ defined by the equation
\begin{equation}
\label{equ:PN}
c_P=k\sqrt{P}\frac{\Theta_2}{\Theta_1}(c_I^2+c_Q^2),
\end{equation}
where $c_I$, $c_Q$, and $c_P$ are the three axes of Cartesian coordinate system of the I-Q-P space. By changing coordinate system, the point $(c_1,c_2, c_3)$ can be represented as $(\tilde{a},l)$ in the PN coordinate system, where $\tilde{a}$ is the nearest point on $\mathcal{U}$ to $(c_1,c_2, c_3)$, and $|l|$ is the distance. In other words, the point $(c_1,c_2, c_3)$ is on the normal line of the paraboloid at the point $\tilde{a}$, and $\tilde{a}$ is the projection of $(c_1,c_2, c_3)$ on $\mathcal{U}$. Specifically, the sign of $l$ is positive when $(c_1,c_2, c_3)$ is above the paraboloid. Otherwise, it is negative. It can be easily verified that both the points $({{{\tilde{X}}}_1}, {{{X}}_2})$ and $({{{\tilde{X}}}'_1}, {{{X}}'_2})$ lie on $\mathcal{U}$.

Using the property of mutual information invariance under a change of coordinates~\cite{BookInfo}, and representing Cartesian coordinate based random variables $({{{\tilde{X}}}_1}, {{{X}}_2})$, $({{{\tilde{X}'}}_1}, {X'_2})$ and $({{{\tilde{Y}}}_1}, {{{Y}}_2})$ in the PN coordinate system as $(\tilde{A}_{{{\tilde{X}}}},L_{\tilde{X}})$,
$(\tilde{A}_{{{\tilde{X}}},\tilde{W}},L_{\tilde{X},\tilde{W}})$ and $(\tilde{A}_{{{\tilde{X}}},\tilde{W}, \tilde{Z}, N},L_{\tilde{X},\tilde{W}, \tilde{Z}, N})$, respectively, we have
\begin{equation}
\label{equ:MIPN}
\mathcal{I}({{{\tilde{X}}}_1}, {{{X}}_2};{{{\tilde{Y}}}_1}, {{{Y}}_2})=\mathcal{I}(\tilde{A}_{{{\tilde{X}}}},L_{\tilde{X}};\tilde{A}_{{{\tilde{X}}},\tilde{W}, \tilde{Z}, N},L_{\tilde{X},\tilde{W}, \tilde{Z}, N}),
\end{equation}
where the noise-related random variables $\tilde{A}_{{{\tilde{X}}},\tilde{W}, \tilde{Z}, N}$ and $L_{\tilde{X},\tilde{W}, \tilde{Z}, N}$ are correlated with the random variable $\tilde{A}_{{{\tilde{X}}}}$.

Recall that both $({{{\tilde{X}}}_1}, {{{X}}_2})$ and $({{{\tilde{X}}}'_1}, {{{X}}'_2})$ lie on $\mathcal{U}$, we have
\begin{equation}
L_{\tilde{X}}=L_{\tilde{X},\tilde{W}}=0,
\end{equation}
and the mutual information is rewritten as
\begin{align}
\label{equ:MIPN1}
&\mathcal{I}({{{\tilde{X}}}_1}, {{{X}}_2};{{{\tilde{Y}}}_1}, {{{Y}}_2})\notag\\
&=\mathcal{I}(\tilde{A}_{{{\tilde{X}}}},L_{\tilde{X}};\tilde{A}_{{{\tilde{X}}},\tilde{W}, \tilde{Z}, N},L_{\tilde{X},\tilde{W}, \tilde{Z}, N})\notag\\
&=\!\mathcal{H}(\tilde{A}_{{{\tilde{X}}},\tilde{W}, \tilde{Z}, N},\!L_{\tilde{X},\! \tilde{W}, \tilde{Z}, N})\!-\!\!\mathcal{H}(\tilde{A}_{{{\tilde{X}}},\! \tilde{W}, \tilde{Z}, N},\!L_{\tilde{X},\tilde{W}, \tilde{Z}, N}|\tilde{A}_{{{\tilde{X}}}}\!)\notag\\
&=\mathcal{H}(\tilde{A}_{{{\tilde{X}}},\tilde{W}, \tilde{Z}, N})-\mathcal{H}(\tilde{A}_{{{\tilde{X}}},\tilde{W}, \tilde{Z}, N}|\tilde{A}_{{{\tilde{X}}}}\!)\notag\\
&\!+\!\!
\left(\!\mathcal{H}(\!L_{\tilde{X},\tilde{W}, \tilde{Z}, N}|\tilde{A}_{{{\tilde{X}}},\tilde{W}, \tilde{Z}, N}\!)\!\!-\!\!\mathcal{H}(\!L_{\tilde{X},\tilde{W}, \tilde{Z}, N}|\tilde{A}_{{{\tilde{X}}}},\! \tilde{A}_{{{\tilde{X}}},\tilde{W}, \tilde{Z}, N}\!)\!\right).
\end{align}

Due to the fact that the expectations $\mathbb{E}(\tilde{Z}) =(0,0)$, $\mathbb{E}(N) =0$  and the variances $\textrm{Var}(\frac{{{\tilde{Z}}}}{\sqrt{P}})$ and $\textrm{Var}(k\frac{{{N}}}{\sqrt{P}})\rightarrow 0$ as $P\rightarrow \infty$, it is easy to see that the random variable $\tilde{A}_{{{\tilde{X}}},\tilde{W}, \tilde{Z}, N}$ converges in probability towards $\tilde{A}_{{{\tilde{X}}},\tilde{W}}$, where $\tilde{A}_{{{\tilde{X}}},\tilde{W}}$ is the projection of $({{{\tilde{X}}}'_1}, {{{X}}'_2})$ on $\mathcal{U}$. Thus, the entropy $\mathcal{H}(\tilde{A}_{{{\tilde{X}}},\tilde{W}, \tilde{Z}, N})$ can be approximated by $\mathcal{H}(\tilde{A}_{{{\tilde{X}}},\tilde{W}})$ when $P$ is large.

Furthermore, from the definition of conditional mutual information~\cite{BookInfo}, it follows that
\begin{align}
&\mathcal{I}(L_{\tilde{X},\tilde{W},\tilde{Z},N};\tilde{A}_{\tilde{X}} \vert \tilde{A}_{\tilde{X},\tilde{W},\tilde{Z},N})\notag\\
&=\!\mathcal{H}(L_{\tilde{X},\tilde{W}, \tilde{Z}, N}|\tilde{A}_{{{\tilde{X}}},\tilde{W}, \tilde{Z}, N}\!)
\!-\!\!
\mathcal{H}(L_{\tilde{X},\tilde{W}, \tilde{Z}, N}|\tilde{A}_{{{\tilde{X}}}},\! \tilde{A}_{{{\tilde{X}}},\tilde{W}, \tilde{Z}, N}\!),
\end{align}
which indicates the additional amount of information of $\tilde{A}_{\tilde{X}}$ by knowing $L_{\tilde{X},\tilde{W}, \tilde{Z}, N}$ when $\tilde{A}_{\tilde{X},\tilde{W}, \tilde{Z}, N}$ is given.
Again, by using the fact that the expectation $\mathbb{E}(\tilde{Z}) =(0,0)$, $\mathbb{E}(N) =0$ and the variances
$\textrm{Var}(\frac{{{\tilde{Z}}}}{\sqrt{P}})$ and $\textrm{Var}(k\frac{{{N}}}{\sqrt{P}})\rightarrow 0$ as $P\rightarrow \infty$, $L_{\tilde{X},\tilde{W}, \tilde{Z}, N}$ converges to the constant $L_{\tilde{X},\tilde{W}}=0$ in probability, when $P\rightarrow \infty$.
Therefore, as $L_{\tilde{X},\tilde{W}, \tilde{Z}, N}$ approaches to zero, the additional amount of information it can provide about $\tilde{A}_{\tilde{X}}$, i.e., $\mathcal{I}(L_{\tilde{X},\tilde{W},\tilde{Z},N};\tilde{A}_{\tilde{X}} \vert \tilde{A}_{\tilde{X},\tilde{W},\tilde{Z},N})$, approaches to zero as well.

Thus, the asymptotic mutual information in \eqnref{equ:MIPN1} can be rewritten as
\begin{align}
\label{equ:MIPN2}
&\mathcal{I}({{{\tilde{X}}}_1}, {{{X}}_2};{{{\tilde{Y}}}_1}, {{{Y}}_2}) =  \mathcal{H}(\tilde{A}_{{{\tilde{X}}},\tilde{W}})-\mathcal{H}(\tilde{A}_{{{\tilde{X}}},\tilde{W}, \tilde{Z}, N}|\tilde{A}_{{{\tilde{X}}}}).
\end{align}

Then, we calculate  $\mathcal{H}(\tilde{A}_{{{\tilde{X}}},\tilde{W}})$ and $\mathcal{H}(\tilde{A}_{{{\tilde{X}}},\tilde{W}, \tilde{Z}, N}|\tilde{A}_{{{\tilde{X}}}})$ in the sequel.

1) $\mathcal{H}(\tilde{A}_{{{\tilde{X}}},\tilde{W}})$: Since $\tilde{X}$ and $(\tilde{X}+\tilde{W})$ are zero-mean complex Gaussian random variables but with different variances, the analysis of $\mathcal{H}(\tilde{A}_{{{\tilde{X}}},\tilde{W}})$ is similar to that of $\mathcal{H}(\tilde{A}_{{\tilde{X}}})$ in~\cite{Liu2017A}. $\mathcal{H}(\tilde{A}_{{{\tilde{X}}},\tilde{W}})$ can be calculated by replacing $f_{\tilde{X}}(\tilde{x})$ in~\cite[Appendix A]{Liu2017A} with
\begin{equation}
\label{equ:pdfx}
f_{\tilde{X}}(\tilde{x})=\frac{1}{\pi\zeta^2}\exp\left( -\frac{|\tilde{x}|^2}{\zeta^2}\right),
\end{equation}
where $\zeta^2=1+\frac{\sigma_{\textrm{A}}^2}{P}$. Following the same steps, we have
\begin{align}
\label{equ:MIAxw}
\mathcal{H}(\tilde{A}_{{{\tilde{X}}},\tilde{W}})=&\log_2(\pi e \Theta_1\zeta^2)\notag\\
&+\frac{1}{2\ln2}\exp\left(\frac{\Theta_1}{4k^2P\Theta_2\zeta^2}\right)\operatorname{Ei}\left(\frac{\Theta_1}{4k^2P\Theta_2\zeta^2}\right),
\end{align}

2) Asymptotic $\mathcal{H}(\tilde{A}_{{{\tilde{X}}},\tilde{W}, \tilde{Z}, N}|\tilde{A}_{{{\tilde{X}}}})$: The conditional entropy can be written as
\begin{align}
\label{equ:MIAxwzn}
\mathcal{H}\big(\tilde{A}_{{{\tilde{X}}},\tilde{W}, \tilde{Z}, N}|\tilde{A}_{{{\tilde{X}}}}\big)&=\mathbb{E}_{\tilde{A}_{{\tilde{X}}}}[\mathcal{H}(\tilde{A}_{{{\tilde{X}}},\tilde{W}, \tilde{Z}, N}|\tilde{A}_{{{\tilde{X}}}}=\tilde{a}_{\tilde{X}})]\notag\\
&=\mathbb{E}_{\tilde{X}}[\mathcal{H}(\tilde{A}_{{{\tilde{X}}},\tilde{W}, \tilde{Z}, N}|{{{\tilde{X}}}})].
\end{align}

Since the random variables $\tilde{X}$, $\tilde{W}$, and $(\tilde{Z},N)$ can be treated as circularly symmetric Gaussian random vectors, the entropy does not reply on the phase of the complex Gaussian random variable $\tilde{X}$ when $\tilde{X}$ is given. Therefore, in what follows, we note that $\tilde{X}=(X_I,0)$ without loss of generality, where $X_I\geq 0$ and $X^2_I$ follows the standard exponential distribution, and thus $X_I$ follows a Rayleigh distribution.

From \eqnref{equ:Variables1}, \eqnref{equ:Variables2}, \eqnref{equ:Variables3}, and \eqnref{equ:Variables4}, we have
\begin{equation}
\label{equ:Variables5}
{{{\tilde{X}}}'_1} = {{\tilde{X}}}_1+\sqrt{\Theta_1}\frac{{{\tilde{W}}}}{\sqrt{P}},
\end{equation}
\begin{align}
\label{equ:Variables6}
{{X}'_2}& = {{{X}}}_2+2\sqrt{\Theta_2}k\sqrt{P}\left(X_I\frac{{{{W}_I}}}{\sqrt{P}}\right)\!+\! \sqrt{\Theta_2}k\sqrt{P}\biggl|\frac{\tilde{W}}{\sqrt{P}}\biggr|^2,\\
\label{equ:Variables7}
&\approx {{{X}}}_2+2\sqrt{\Theta_2}k\sqrt{P}\left(X_I\frac{{{\tilde{W}}}}{\sqrt{P}}\right),
\end{align}
where $W_I$ and $W_Q$ are the real and imaginary parts of $\tilde{W}$, respectively. Then the approximation \eqnref{equ:Variables7} is valid when $P$ is sufficiently large, i.e., the antenna noise term $\frac{\tilde{W}}{\sqrt{P}}$ is very small.

From \eqnref{equ:Variables5} and \eqnref{equ:Variables7}, the point $({{{\tilde{X}}}'_1},{{X}'_2})$ lies on the paraboloid $\mathcal{U}$ in the I-Q-P space can be approximated as
\begin{align}
\label{equ:Variables8}
&({{{\tilde{X}}}'_1},{{X}'_2})\approx ({{{\tilde{X}}}_1},{{X}_2})\!+\!\biggl(\sqrt{\Theta_1}\frac{{{\tilde{W}}}}{\sqrt{P}},2\sqrt{\Theta_2}k\sqrt{P}X_I\frac{{{{W}_I}}}{\sqrt{P}}\biggr)\notag\\
&=({{{\tilde{X}}}_1},{{X}_2})\!+\!\sqrt{\Theta_1+4\Theta_2Pk^2X_I^2}\frac{{W_I}}{\sqrt{P}}\textbf{i}_{IP}\!+\!\sqrt{\Theta_1}\frac{{W_Q}}{\sqrt{P}}\textbf{i}_Q,
\end{align}
where $\textbf{i}_{IP}$ and $\textbf{i}_{Q}$ are mutually orthogonal unit vectors and
\begin{align}
\label{equ:unvec}
&\textbf{i}_{IP}\triangleq \biggl(\frac{\sqrt{\Theta_1}}{\sqrt{\Theta_1+4\Theta_2Pk^2X_I^2}},0,\frac{2\sqrt{\Theta_2}k\sqrt{P}X_I}{\sqrt{\Theta_1+4\Theta_2Pk^2X_I^2}}\biggr),\notag\\
&\textbf{i}_{Q}\triangleq (0,1,0).
\end{align}

It is easy to verify that the approximated point $({{{\tilde{X}}}'_1},{{X}'_2})$ in \eqref{equ:Variables8} lies on the tangent plane of $\mathcal{U}$ of the point $({{{\tilde{X}}}_1},{{X}_2})$, $\forall W_I,W_Q\in \mathbb{R}$, for large $P$. Taking \eqnref{equ:Variables8} into \eqnref{equ:spre}, we have
\begin{align}
\label{equ:sprech}
&(\tilde{Y}_1,Y_2)=({{{\tilde{X}}}'_1},{{X}'_2})+\biggl(\frac{\tilde{Z}}{\sqrt{P}},\frac{kN}{\sqrt{P}}\biggr)\notag\\
&\approx ({{{\tilde{X}}}_1},{{X}_2})+\left(\sqrt{\Theta_1+4\Theta_2Pk^2X_I^2}\frac{{W_I}}{\sqrt{P}}+\frac{Z_1}{\sqrt{P}} \right)\textbf{i}_{IP}\notag\\
&\,\,\,\,\,\,\,+\left( \sqrt{\Theta_1}\frac{{W_Q}}{\sqrt{P}}+\frac{Z_2}{\sqrt{P}}\right)\textbf{i}_Q+\frac{Z_3}{\sqrt{P}}\textbf{i}_{IQP},
\end{align}
where $\textbf{i}_{IQP}$ is a unit vector that is orthogonal to $\textbf{i}_{IP}$ and $\textbf{i}_Q$, and $\big(\frac{\tilde{Z}}{\sqrt{P}},\frac{kN}{\sqrt{P}}\big)\triangleq\frac{Z_1}{\sqrt{P}}\textbf{i}_{IP}+\frac{Z_2}{\sqrt{P}}\textbf{i}_{Q}+\frac{Z_3}{\sqrt{P}}\textbf{i}_{IQP}$. Since $\big(\frac{\tilde{Z}}{\sqrt{P}},\frac{kN}{\sqrt{P}}\big)$ is a 3D circularly symmetric Gaussian random vector and $\textbf{i}_{IP}$, $\textbf{i}_{Q}$, and $\textbf{i}_{IQP}$ are mutually orthogonal unit vectors, $Z_1$, $Z_2$, and $Z_3$ are i.i.d. zero-mean real Gaussian random variables with the variance $\frac{\sigma_{\textrm{cov}}^2}{2}$ in \eqref{equ:sprech}.

When the noise terms are sufficiently small, $(\tilde{Y}_1,Y_2)$ converges to $({{{\tilde{X}}}_1},{{X}_2})$ on $\mathcal{U}$ in probability. Thus, the projection of $(\tilde{Y}_1,Y_2)$ on $\mathcal{U}$, i.e., $\tilde{A}_{{{\tilde{X}}},\tilde{W}, \tilde{Z}, N}$, converges to the projection on the talent plane of the point $({{{\tilde{X}}}_1},{{X}_2})$ on $\mathcal{U}$, named as $\mathcal{S}$. Therefore, from \eqnref{equ:sprech}, $\tilde{A}_{{{\tilde{X}}},\tilde{W}, \tilde{Z}, N}$ can be approximated as
\begin{align}
\label{equ:Axwznapp}
\tilde{A}_{{{\tilde{X}}},\tilde{W}, \tilde{Z}, N}\approx&({{{\tilde{X}}}_1},{{X}_2})+\left( \sqrt{\Theta_1}\frac{{W_Q}}{\sqrt{P}}+\frac{Z_2}{\sqrt{P}}\right)\textbf{i}_Q \notag\\
& +\left(\sqrt{\Theta_1+4\Theta_2Pk^2X_I^2}\frac{{W_I}}{\sqrt{P}}+\frac{Z_1}{\sqrt{P}} \right)\textbf{i}_{IP}.
\end{align}

Given $\tilde{X}$, since $W_I$, $W_Q$, $Z_1$ and $Z_2$ are zero-mean Gaussian random variables and are independent with each other, the approximated $\tilde{A}_{{{\tilde{X}}},\tilde{W}, \tilde{Z}, N}$ in \eqref{equ:Axwznapp} is a complex Gaussian random variable on $\mathcal{S}$ with the covariance matrix
\begin{align}
\label{equ:Var}
\Sigma=\left[\begin{array}{cccc}
    \frac{\big(\Theta_1+4\Theta_2Pk^2X_I^2\big)\sigma_{\textrm{A}}^2+\sigma_{\textrm{cov}}^2}{2P} & \!\!\! 0 \!\!\!\\
    \!\!\!0 \!\!\! &    \frac{\Theta_1\sigma_{\textrm{A}}^2+\sigma_{\textrm{cov}}^2}{2P}\\
\end{array}\right].
\end{align}

Thus, from \eqnref{equ:Var}, the conditional entropy is approximated~as
\begin{align}
\label{equ:MIAxwznapp}
&\mathcal{H}(\tilde{A}_{{{\tilde{X}}},\tilde{W}, \tilde{Z}, N}|X_I=x)\notag\\
&=  \log_2\pi e+\frac{1}{2}\log_2\left( \frac{\Theta_1\sigma_{\textrm{A}}^2+\sigma_{\textrm{cov}}^2}{P}\right)\notag\\
&\,\,\,\,\,\,\,+\frac{1}{2}\log_2\left(\frac{(\Theta_1+4\Theta_2Pk^2x^2)\sigma_{\textrm{A}}^2+\sigma_{\textrm{cov}}^2}{P} \right).
\end{align}

Therefore, taking \eqnref{equ:MIAxwznapp} into \eqnref{equ:MIAxwzn}, the approximated conditional entropy is
\begin{align}
\label{equ:MIAxwznend}
&\mathcal{H}(\tilde{A}_{{{\tilde{X}}},\tilde{W}, \tilde{Z}, N}|\tilde{A}_{{{\tilde{X}}}})=\mathcal{H}(\tilde{A}_{{{\tilde{X}}},\tilde{W}, \tilde{Z}, N}|X_I)\notag\\
&= \int_{0}^{\infty} \biggl( \log_2\pi e+\frac{1}{2}\log_2\biggl( \frac{\Theta_1\sigma_{\textrm{A}}^2+\sigma_{\textrm{cov}}^2}{P}\biggr)\notag\\
&\,\,\,\,\,\,+\frac{1}{2}\log_2\biggl(\frac{(\Theta_1\!+\!4\Theta_2Pk^2 r)\sigma_{\textrm{A}}^2\!+\!\sigma_{\textrm{cov}}^2}{P} \biggr)\biggr) e^{-r}\textrm{d}r\notag\\
&=\log_2\pi e+\log_2\biggl( \frac{\Theta_1\sigma_{\textrm{A}}^2+\sigma_{\textrm{cov}}^2}{P}\biggr)\notag\\
&\,\,\ \,+\frac{1}{2}\biggl( \frac{1}{\ln2}\exp\biggl( \frac{\Theta_1\sigma_{\textrm{A}}^2+\sigma_{\textrm{cov}}^2}{4\Theta_2Pk^2\sigma_{\textrm{A}}^2}\biggr)\operatorname{Ei}\biggr( \frac{\Theta_1\sigma_{\textrm{A}}^2+\sigma_{\textrm{cov}}^2}{4\Theta_2Pk^2\sigma_{\textrm{A}}^2}\biggr)\biggr).
\end{align}

3) Asymptotic $\mathcal{I}(\sqrt{P}{{\tilde{X}}};{{{\tilde{Y}}}_1}, {{{Y}}_2})$:
Taking \eqnref{equ:MIAxw} and \eqnref{equ:MIAxwznend} into \eqnref{equ:MIPN2}, the asymptotic mutual information can be obtained as
\begin{align}
\label{equ:MIend}
&\mathcal{I}({{{\tilde{X}}}_1}, {{{X}}_2};{{{\tilde{Y}}}_1}, {{{Y}}_2})\notag\\
&=\log_2\biggl( \frac{\Theta_1\zeta^2P}{\Theta_1\sigma_{\textrm{A}}^2+\sigma_{\textrm{cov}}^2}\biggr)\notag\\
&\,\,\,\,\,\, +\frac{1}{2\ln2}\biggl(\exp\biggl(\frac{\Theta_1}{4k^2P\Theta_2\zeta^2}\biggr)\operatorname{Ei}\biggl(\frac{\Theta_1}{4k^2P\Theta_2\zeta^2}\biggr)\notag\\
&\,\,\,\,\,\,-\exp\biggl( \frac{\Theta_1\sigma_{\textrm{A}}^2+\sigma_{\textrm{cov}}^2}{4\Theta_2Pk^2\sigma_{\textrm{A}}^2}\biggr)\operatorname{Ei}\biggl( \frac{\Theta_1\sigma_{\textrm{A}}^2+\sigma_{\textrm{cov}}^2}{4\Theta_2Pk^2\sigma_{\textrm{A}}^2}\biggr)\biggr).
\end{align}
Taking $\zeta^2=1+\frac{\sigma_{\textrm{A}}^2}{P}$ and $k^2=\frac{\sigma_{\textrm{cov}}^2}{2\sigma_{\textrm{rec}}^2}$ into \eqnref{equ:MIend}, we have
\begin{align}
\label{equ:MIhighSNRapp}
&\mathcal{I}(\sqrt{P}{{\tilde{X}}};{{{\tilde{Y}}}_1},{{{Y}}_2})= \log_2\left(\frac{\Theta_1(P+\sigma_{\textrm{{A}}}^2)}{\Theta_1\sigma_{\textrm{{A}}}^2+\sigma_{\textrm{{cov}}}^2}\right)\notag\\
&+\frac{1}{2\ln2}\biggl( \exp\biggl( \frac{\Theta_1\sigma_{\textrm{{rec}}}^2}{2\Theta_2\sigma_{\textrm{{cov}}}^2(P+\sigma_{\textrm{{A}}}^2)}\biggr)\operatorname{Ei}\biggl( \frac{\Theta_1\sigma_{\textrm{{rec}}}^2}{2\Theta_2\sigma_{\textrm{{cov}}}^2(P+\sigma_{\textrm{{A}}}^2)}\biggr)\notag\\
&-\exp\biggl( \frac{(\Theta_1\sigma_{\textrm{{A}}}^2+\sigma_{\textrm{{cov}}}^2)\sigma_{\textrm{{rec}}}^2}{2\Theta_2P\sigma_{\textrm{{A}}}^2\sigma_{\textrm{{cov}}}^2}\biggr)\operatorname{Ei}\biggl( \frac{(\Theta_1\sigma_{\textrm{{A}}}^2+\sigma_{\textrm{{cov}}}^2)\sigma_{\textrm{{rec}}}^2}{2\Theta_2P\sigma_{\textrm{{A}}}^2\sigma_{\textrm{{cov}}}^2}\biggr)\biggr).
\end{align}

Taking $\Theta_1\triangleq \rho $ and $\Theta_2\triangleq (1-\rho)^2 $ into \eqnref{equ:MIhighSNRapp}, and replacing
$P$ and $\sqrt{P}$ with $P|\tilde{h}|^2$ and $\sqrt{P}|\tilde{h}|$, respectively,
\eqnref{equ:MIhighSNR} is obtained.

\section{Proof of Proposition 2}
We first derive an asymptotic expression of the mutual information of the splitting receiver with $\rho\in(0,1)$. Based on the power series expansion of the exponential integral function \cite{abramowitz1965handbook}
\begin{equation}
\label{equ:EI}
\operatorname{Ei}(x)=-\gamma-\ln x-\sum_{n=1}^{\infty}\frac{(-x)^n}{nn!},x>0,
\end{equation}
where $\gamma\approx 0.5772$ is Euler's constant, we have
\begin{equation}
\label{equ:EI1}
\mathop{\lim} \limits_{x_1,x_2 \rightarrow 0} \exp(x_1)\operatorname{Ei}(x_1)-\exp(x_2)\operatorname{Ei}(x_2)=\ln\left(\frac{x_2}{x_1}\right).
\end{equation}
Substituting \eqnref{equ:EI1} into \eqnref{equ:MIhighSNRapp}, \eqnref{equ:MIhighSNRapp} is simplified as
\begin{equation}
\label{equ:MIhighSNR1}
\mathop{\lim} \limits_{P  \rightarrow\infty} \mathcal{I}(\sqrt{P}\vert \tilde{h} \vert{{\tilde{X}}};{{{\tilde{Y}}}_1},{{{Y}}_2})
= \log_2\left( \frac{P|\tilde{h}|^2}{\sqrt{\sigma_{\textrm{{A}}}^2+\frac{\sigma_{\textrm{{cov}}}^2}{\rho}}\sqrt{\sigma_{\textrm{{A}}}^2}}\right).
\end{equation}

In order to calculate the mutual information performance gain as seen in Definition~\ref{define:MIde}, we need to obtain the asymptotic mutual information of the non-coherent PD receiver ($\rho=0$) and CD receiver ($\rho=1$), respectively. The mutual information of the CD receiver has a well-known closed-form expression in \eqnref{equ:MICDreceiver}. However, the mutual information of the PD receiver in \eqnref{equ:MIPDreceiver} has multiple integrals to evaluate, which is difficult to simplify, and thus we resort to a different approach. That is, we find an upper bound on the mutual information of the PD receiver and show that it is smaller than the mutual information achieved by the CD receiver in the high SNR regime.

An upper bound on the mutual information achieved by the PD receiver can be obtained by ignoring the rectifier noise $N$. Setting $N=0$, the received signal in \eqnref{equ:PDreceiverch} with $\rho=0$ is simplified as
\begin{equation}
\label{equ:PDreceiverchsimrho0}
{{{Y}}_2} =\big|\sqrt{P}|\tilde{h}|{{\tilde{X}}}+{{\tilde{W}}}\big|^2.
\end{equation}

The mutual information for the channel model above has been studied in the literature. The optimal input distribution is discrete and possesses an infinite number of mass points~\cite{katz2004capacity}. Specifically, an upper-bound of the mutual information is given by~\cite{zhou2013wireless,katz2004capacity}
\begin{align}
\label{equ:NOcapupp}
&\mathcal{I}(\sqrt{P}|\tilde{h}|{{\tilde{X}}};{{{Y}}_2})\notag\\
&\leq \frac{1}{2}\log_2\left(1+\frac{P|\tilde{h}|^2}{2\sigma_{\textrm{A}}^2}\right)+\frac{1}{2}\left(\log_2\frac{2\pi}{e}-\gamma\log_2e\right).
\end{align}

Based on \eqnref{equ:NOcapupp}, the upper bound on mutual information of the PD receiver scales with $P$ as $\frac{1}{2}\log_2 P$. On the other hand, the achievable mutual information of the CD receiver in \eqnref{equ:MICDreceiver} scales with $P$ as $\log_2 P$. Therefore, it can be easily shown that the CD receiver achieves much higher mutual information than the PD receiver as $P\rightarrow\infty$, i.e.,
$\mathcal{I}(\sqrt{P}|\tilde{h}|{{\tilde{X}}};{{{\tilde{Y}}}_1},{{{Y}}_2})|_{\rho=0}\ll \mathcal{I}(\sqrt{P}|\tilde{h}|{{\tilde{X}}};{{{\tilde{Y}}}_1},{{{Y}}_2})|_{\rho=1}$.
Thus, we have
\begin{align}
\label{equ:MIgain1}
G_{\textrm{MI}}= \sup\{\mathcal{I}(\sqrt{P}|\tilde{h}|{{\tilde{X}}};{{{\tilde{Y}}}_1},&{{{Y}}_2}): \rho\in (0,1)\}\notag\\
&-\mathcal{I}(\sqrt{P}|\tilde{h}|{{\tilde{X}}};{{{\tilde{Y}}}_1},{{{Y}}_2})|_{\rho=1}.
\end{align}

According to \eqnref{equ:MIhighSNR1} and \eqnref{equ:MICDreceiver}, the performance improvement of the splitting receiver is expressed as
\begin{align}
\label{equ:MIgain2a}
\mathop{\lim} \limits_{P\rightarrow\infty}G_{\textrm{MI}}\!&= \! \mathcal{I}(\!\sqrt{P}|\tilde{h}|{{\tilde{X}}};{{{\tilde{Y}}}_1},\!{{{Y}}_2}\!)|_{\rho\rightarrow1}\!-\!\mathcal{I}(\!\sqrt{P}|\tilde{h}|{{\tilde{X}}};{{{\tilde{Y}}}_1},\!{{{Y}}_2}\!)|_{\rho=1}\notag\\
&\!=\!\log_2\left( \frac{P|\tilde{h}|^2}{\sqrt{\sigma_{\textrm{{A}}}^2\!+\!{\sigma_{\textrm{{cov}}}^2}}\sqrt{\sigma_{\textrm{{A}}}^2}}\right)\!\!-\!\log_2\left(1\!+\!\frac{P|\tilde{h}|^2}{\sigma_{\textrm{cov}}^2\!+\!\sigma_{\textrm{A}}^2}\right)\notag\\
&=\frac{1}{2}\log_2\left(1+\frac{\sigma_{\textrm{{cov}}}^2}{\sigma_{\textrm{{A}}}^2}\right).
\end{align}

Using \eqnref{equ:MIgain2a} and \eqnref{equ:MICDreceiver}, \eqnref{equ:MIgain2} and \eqnref{equ:MIgain3} are obtained directly.

\section{Proof of Proposition 3}
For a candidate transmit symbol $\tilde{X}$, we define two terms:
$\tilde{T}_1\triangleq{{{\tilde{y}}}_1}
-\sqrt{\rho}\sqrt{P}|\tilde{h}|{{\tilde{X}}}$ and $T_2\triangleq{{{y}}_2} - (1-\rho)\sqrt{P}|\tilde{h}|^2|{{\tilde{X}}}|^2$.
Based on \eqnref{equ:CDreceiver} and \eqnref{equ:PDreceiverch2C1}, the conditional PDF $\hat{f}_{\tilde{Y}_{1}, Y_2}\big(\tilde{y}_{1}, y_2|\tilde{X}\big)$ is given by
\begin{align}
\label{equ:3D2pdf1}
&\hat{f}_{\tilde{Y}_{1}, Y_2}\big(\tilde{y}_1,y_2|\tilde{X}\big) \notag\\
&=\int_{{w}_r}\int_{{w}_i}\hat{f}_{\tilde{Y}_{1},
Y_2}\big(\tilde{y}_{1},y_2|{w}_r,{w}_i,\tilde{X}\big)f_{{\tilde{W}}}\big({w}_r,{w}_i\big)\,\mathrm{d}{w}_i\mathrm{d}{w}_r\notag\\
&=\int_{{w}_r}\int_{{w}_i}\hat{f}_{Y_2}\big(y_2|{w}_r,{w}_i,\tilde{X}\big)f_{\tilde{Y}_{1}}\big(\tilde{y}_{1}|{w}_r,{w}_i,\tilde{X}\big)\notag\\
&\,\,\,\,\,\,\,f_{{\tilde{W}}}\big({w}_r,{w}_i\big)\,\mathrm{d}{w}_i\mathrm{d}{w}_r\notag\\
&=\int_{{w}_r}\int_{{w}_i}f_{{N}_s}\left(T_2-2|\tilde{h}|(1-\rho)({X}_r{{{w}}}_r+{X}_i{{{w}}}_i)\right)\notag\\
&\,\,\,\,\,\,\,f_{\tilde{Z}}\left(\tilde{T}_1-\sqrt{\rho}{{\tilde{w}}}\right)f_{{\tilde{W}}}\left({w}_r,{w}_i\right)\,\mathrm{d}{w}_i\mathrm{d}{w}_r,
\end{align}
where ${{N}}_s=\frac{{N}}{\sqrt{P}}$ is the scaled version of the rectifier noise, which follows $\mathcal{N} (0, \sigma_{{N}_s}^2)$, and $\sigma_{{N}_s}^2=\frac{\sigma_{\textrm{rec}}^2}{P}$. In addition, $f_{{{{N}_s}}}(\cdot)$ and $f_{{{\tilde{Z}}}}(\cdot)$ denote the PDFs of ${{N}_s}$ and ${\tilde{Z}}$, respectively. Hence, the conditional PDF of $\hat{f}_{\tilde{Y}_{1}, Y_2}\big(\tilde{y}_{1}, y_2|\tilde{X}\big)$ is further derived~as
\begin{align}
\label{equ:3D2pdf2}
&\hat{f}_{\tilde{Y}_{1}, Y_2}\big(\tilde{y}_1,y_2|\tilde{X}\big)=\notag\\\!
&\int_{{w}_r}\int_{{w}_i}\!\frac{1}{\sqrt{2\pi\sigma_{{N}_s}^2}}\!\exp\!\!\left(\!\!-\frac{\big(T_2\!-\!2|\tilde{h}|(1\!-\!\rho)({X}_r{{{w}}}_r\!+\!{X}_i{{{w}}}_i)\big)^2}{2\sigma_{{N}_s}^2}\!\right)\notag\\
&\frac{1}{\pi{\sigma_{\textrm{cov}}^2}}\exp\left(-\frac{\big(T_{1r}-\sqrt{\rho}{{{w}}_r}\big)^2+\big(T_{1i}-\sqrt{\rho}{{w}_i}\big)^2}{{\sigma_{\textrm{cov}}^2}}\right)\notag\\
&\frac{1}{\pi\sigma_{\textrm{A}}^2}\exp\left(-\frac{w_r^2+w_i^2}{\sigma_{\textrm{A}}^2}\right)\,\mathrm{d}{w}_i\mathrm{d}{w}_r.
\end{align}

To solve the double integration, we rewrite \eqnref{equ:3D2pdf2} as follows:
\begin{align}
\label{equ:3D2pdf3}
&\hat{f}_{\tilde{Y}_{1},
Y_2}\big(\tilde{y}_1,y_2|\tilde{X}\big)=\frac{1}{\pi^2\sigma_{\textrm{A}}^2{\sigma_{\textrm{cov}}^2}\sqrt{2\pi\sigma_{{N}_s}^2}}\int_{{w}_r}\notag\\
&\exp\biggl(\!\!-\frac{4|\tilde{h}|^2(1-\rho)^2|{X}_r|^2w_r^2-4|\tilde{h}|(1-\rho)T_2{X}_rw_r+|T_2|^2}{2\sigma_{{N}_s}^2}\notag\\
&-\frac{|T_{1r}|^2+|T_{1i}|^2+\rho
w_r^2-2T_{1r}\sqrt{\rho}{w}_r}{{\sigma_{\textrm{cov}}^2}}-\frac{w_r^2}{\sigma_{\textrm{A}}^2}\biggr)\notag\\
&\biggl(
\int_{{w}_i}\!\exp\!\biggl(\!-\frac{4|\tilde{h}|^2(1-\rho)^2|{X}_i|^2w_i^2-4|\tilde{h}|(1-\rho)T_2{X}_iw_i}{2\sigma_{{N}_s}^2}+\notag\\
&\frac{8|\tilde{h}|^2(1\!-\!\rho)^2{X}_i{X}_rw_iw_r}{2\sigma_{{N}_s}^2}\!-\!\frac{\rho
w_i^2\!\!-\!\!2\sqrt{\rho}T_{1i}{w}_i}{{\sigma_{\textrm{cov}}^2}}\!\!-\!\!\frac{w_i^2}{\sigma_{\textrm{A}}^2}\!\biggr)\mathrm{d}{w}_i
\biggr)\mathrm{d}{w}_r,
\end{align}
where $\tilde{T}_1=T_{1r}+jT_{1i}$. Specifically, $T_{1r}\triangleq{{{{y}}}_{1r}} -\sqrt{\rho}\sqrt{P}|\tilde{h}|{{{X}}_r}$;
$T_{1i}\triangleq{{{{y}}}_{1i}}
-\sqrt{\rho}\sqrt{P}|\tilde{h}|{{{X}}_i}$.
By computing the double integrals, the closed-form expression of $\hat{f}_{\tilde{Y}_{1}, Y_2}\big(\tilde{y}_{1}, y_2|\tilde{X}\big)$ is obtained in \eqnref{equ:MLsub2opt1}. Thus, the low-complexity detection rule is obtained.
  \end{appendices}


\begin{thebibliography}{10}
	\providecommand{\url}[1]{#1}
	\csname url@samestyle\endcsname
	\providecommand{\newblock}{\relax}
	\providecommand{\bibinfo}[2]{#2}
	\providecommand{\BIBentrySTDinterwordspacing}{\spaceskip=0pt\relax}
	\providecommand{\BIBentryALTinterwordstretchfactor}{4}
	\providecommand{\BIBentryALTinterwordspacing}{\spaceskip=\fontdimen2\font plus
		\BIBentryALTinterwordstretchfactor\fontdimen3\font minus
		\fontdimen4\font\relax}
	\providecommand{\BIBforeignlanguage}[2]{{%
			\expandafter\ifx\csname l@#1\endcsname\relax
			\typeout{** WARNING: IEEEtran.bst: No hyphenation pattern has been}%
			\typeout{** loaded for the language `#1'. Using the pattern for}%
			\typeout{** the default language instead.}%
			\else
			\language=\csname l@#1\endcsname
			\fi
			#2}}
	\providecommand{\BIBdecl}{\relax}
	\BIBdecl
	
	\bibitem{PetarMag}
	F.~Boccardi, R.~W. Heath~Jr., A.~Lozano, T.~L. Marzetta, and P.~Popovski,
	``Five disruptive technology directions for {5G},'' \emph{IEEE Commun. Mag.},
	vol.~52, no.~2, pp. 74--80, Feb. 2014.
	
	\bibitem{larsson2014massive}
	E.~G. Larsson, O.~Edfors, F.~Tufvesson, and T.~L. Marzetta, ``Massive {MIMO}
	for next generation wireless systems,'' \emph{IEEE Commun. Mag.}, vol.~52,
	no.~2, pp. 186--195, Feb. 2014.
	
	\bibitem{proakis2007digital}
	J.~G. Proakis and M.~Salehi, \emph{Digital communications}.\hskip 1em plus
	0.5em minus 0.4em\relax McGraw-hill New York, 2007, vol.~5.
	
	\bibitem{IQ}
	S.~{Lajnef}, N.~{Boulejfen}, A.~{Abdelhafiz}, and F.~M. {Ghannouchi},
	``Two-dimensional cartesian memory polynomial model for nonlinearity and
	{I/Q} imperfection compensation in concurrent dual-band transmitters,''
	\emph{IEEE Trans. Circuits Syst. II, Exp. Briefs 2}, vol.~63, no.~1, pp.
	14--18, Jan. 2016.
	
	\bibitem{lowpower}
	C.-Y. Chu, C.-C. Wei, H.-C. Hsu, S.-H. Feng, and W.-S. Feng, ``A 24 {GH}z
	low-power {CMOS} receiver design,'' in \emph{Proc. IEEE Inter. Symp. Circuits
		Syst. (ISCAS)}, Seattle, WA, USA, May 2008, pp. 980--983.
	
	\bibitem{Mo}
	J.~{Mo} and R.~W. {Heath Jr.}, ``Capacity analysis of one-bit quantized {MIMO}
	systems with transmitter channel state information,'' \emph{IEEE Trans.
		Signal Process.}, vol.~63, no.~20, pp. 5498--5512, Oct. 2015.
	
	\bibitem{Li}
	Y.~{Li}, C.~{Tao}, G.~{Seco-Granados}, A.~{Mezghani}, A.~L. {Swindlehurst}, and
	L.~{Liu}, ``Channel estimation and performance analysis of one-bit massive
	{MIMO} systems,'' \emph{IEEE Trans. Signal Process.}, vol.~65, no.~15, pp.
	4075--4089, Aug. 2017.
	
	\bibitem{Chowdhury2016Scaling}
	M.~{Chowdhury}, A.~{Manolakos}, and A.~{Goldsmith}, ``Scaling laws for
	noncoherent energy-based communications in the {SIMO} {MAC},'' \emph{IEEE
		Trans. Inf. Theory}, vol.~62, no.~4, pp. 1980--1992, Apr. 2016.
	
	\bibitem{Jing2016Design}
	L.~Jing, E.~De~Carvalho, P.~Popovski, and A.~O. Martinez, ``Design and
	performance analysis of noncoherent detection systems with massive receiver
	arrays,'' \emph{IEEE Trans. Signal Process.}, vol.~64, no.~19, pp.
	5000--5010, Oct. 2016.
	
	\bibitem{liu2019next}
	W.~Liu, K.~Huang, X.~Zhou, and S.~Durrani, ``Next generation backscatter
	communication: systems, techniques, and applications,'' \emph{EURASIP J.
		Wireless Commun. Netw.}, vol.~69, pp. 1--11, Mar. 2019.
	
	\bibitem{Liu2017A}
	W.~Liu, X.~Zhou, S.~Durrani, and P.~Popovski, ``A novel receiver design with
	joint coherent and non-coherent processing,'' \emph{IEEE Trans. Commun.},
	vol.~65, no.~8, pp. 3479--3493, Aug. 2017.
	
	\bibitem{Wilkinsons}
	\BIBentryALTinterwordspacing
	microwaves101.com, \emph{Unequal-Split Wilkinsons}, Aug. 2009. [Online].
	Available:
	\url{https://www.microwaves101.com/encyclopedias/unequal-split-wilkinsons}
	\BIBentrySTDinterwordspacing
	
	\bibitem{zhang2013mimo}
	R.~Zhang and C.~K. Ho, ``{MIMO} broadcasting for simultaneous wireless
	information and power transfer,'' \emph{IEEE Trans. Wireless Commun.},
	vol.~12, no.~5, pp. 1989--2001, May 2013.
	
	\bibitem{wu2010analytical}
	Y.~Wu, Y.~Liu, Q.~Xue, S.~Li, and C.~Yu, ``Analytical design method of multiway
	dual-band planar power dividers with arbitrary power division,'' \emph{IEEE
		Trans. Microw. Theory Techn.}, vol.~58, no.~12, pp. 3832--3841, Dec. 2010.
	
	\bibitem{PhysRevLett}
	R.~F. Voss, ``Linearity of $1/f$ noise mechanisms,'' \emph{Phys. Rev. Lett.},
	vol.~40, pp. 913--916, Apr. 1978.
	
	\bibitem{zhou2013wireless}
	X.~Zhou, R.~Zhang, and C.~K. Ho, ``Wireless information and power transfer:
	Architecture design and rate-energy tradeoff,'' \emph{IEEE Trans. Commun.},
	vol.~61, no.~11, pp. 4754--4767, Nov. 2013.
	
	\bibitem{lapidoth2009capacity}
	A.~Lapidoth, S.~M. Moser, and M.~A. Wigger, ``On the capacity of free-space
	optical intensity channels,'' \emph{IEEE Trans. Inf. Theory}, vol.~55,
	no.~10, pp. 4449--4461, Oct. 2009.
	
	\bibitem{infrared1997}
	J.~M. {Kahn} and J.~R. {Barry}, ``Wireless infrared communications,''
	\emph{Proc. IEEE}, vol.~85, no.~2, pp. 265--298, Feb. 1997.
	
	\bibitem{gu2006rf}
	Q.~Gu, \emph{{RF} system design of transceivers for wireless
		communications}.\hskip 1em plus 0.5em minus 0.4em\relax Springer Science \&
	Business Media, 2006.
	
	\bibitem{lin2013achieving}
	C.~Lin and B.-S. Chen, ``Achieving pareto optimal power tracking control for
	interference limited wireless systems via multi-objective $h_2/h_{\infty}$
	optimization,'' \emph{IEEE Trans. Wireless Commun.}, vol.~12, no.~12, pp.
	6154--6165, Dec. 2013.
	
	\bibitem{yang2018dense}
	B.~Yang, G.~Mao, M.~Ding, X.~Ge, and X.~Tao, ``Dense small cell networks: From
	noise-limited to dense interference-limited,'' \emph{IEEE Trans. Veh.
		Technol.}, vol.~67, no.~5, pp. 4262--4277, May 2018.
	
	\bibitem{perez2018signal}
	A.~I. {Perez-Neira}, M.~A. {Vazquez}, M.~R.~B. {Shankar}, S.~{Maleki}, and
	S.~{Chatzinotas}, ``Signal processing for high-throughput satellites:
	Challenges in new interference-limited scenarios,'' \emph{IEEE Signal
		Process. Mag.}, vol.~36, no.~4, pp. 112--131, July 2019.
	
	\bibitem{razavi1998rf}
	B.~Razavi and R.~Behzad, \emph{RF microelectronics}.\hskip 1em plus 0.5em minus
	0.4em\relax Prentice hall New Jersey, 1998, vol.~1.
	
	\bibitem{hendriks2001high}
	P.~Hendriks, R.~Schreier, and J.~DiPilato, ``High performance narrowband
	receiver design simplified by {IF} digitizing subsystem in {LQFP},''
	\emph{Analog Dialogue}, vol.~35, p.~3, 2001.
	
	\bibitem{loy1999understanding}
	M.~Loy, ``Understanding and enhancing sensitivity in receivers for wireless
	applications,'' \emph{Texas Instruments, Technical Brief}, pp. 1--77, 1999.
	
	\bibitem{clerckx2018fundamentals}
	B.~Clerckx, R.~Zhang, R.~Schober, D.~W.~K. Ng, D.~I. Kim, and H.~V. Poor,
	``Fundamentals of wireless information and power transfer: From {RF} energy
	harvester models to signal and system designs,'' \emph{IEEE J. Sel. Areas
		Commun.}, vol.~37, no.~1, pp. 4--33, Jan. 2019.
	
	\bibitem{flicker}
	J.~L. Hesler and T.~W. Crowe, ``Responsivity and noise measurements of
	zero-bias {S}chottky diode detectors,'' in \emph{Proc. Intl. Symp. Space
		Terahertz Techn.(ISSTT)}, Pasadena, CA, Mar. 2007, pp. 89--92.
	
	\bibitem{haney2011practical}
	S.~Haney, \emph{Practical applications and properties of the Exponentially
		Modified Gaussian ({EMG}) distribution}.\hskip 1em plus 0.5em minus
	0.4em\relax Ph.D. dissertation, Drexel Univ., Philadelphia, PA, USA, 2011.
	
	\bibitem{grushka1972characterization}
	E.~Grushka, ``Characterization of exponentially modified {G}aussian peaks in
	chromatography,'' \emph{Anal. Chem.}, vol.~44, no.~11, pp. 1733--1738, 1972.
	
	\bibitem{siegel1979noncentral}
	A.~F. Siegel, ``The noncentral chi-squared distribution with zero degrees of
	freedom and testing for uniformity,'' \emph{Biometrika}, vol.~66, no.~2, pp.
	381--386, 1979.
	
	\bibitem{michalowicz2013handbook}
	J.~V. Michalowicz, J.~M. Nichols, and F.~Bucholtz, \emph{Handbook of
		differential entropy}.\hskip 1em plus 0.5em minus 0.4em\relax Chapman and
	Hall/CRC, 2013.
	
	\bibitem{katz2004capacity}
	M.~Katz and S.~Shamai, ``On the capacity-achieving distribution of the
	discrete-time noncoherent and partially coherent {AWGN} channels,''
	\emph{IEEE Trans. Inf. Theory}, vol.~50, no.~10, pp. 2257--2270, Oct. 2004.
	
	\bibitem{BookInfo}
	T.~Cover and J.~Thomas, \emph{Elements of Information Theory}.\hskip 1em plus
	0.5em minus 0.4em\relax Wiley, 2006.
	
	\bibitem{etsi2014302}
	E.~ETSI, ``302 307-2 (v1.1.1): “{D}igital video broadcasting ({DVB}): Second
	generation framing structure, channel coding and modulation systems for
	broadcasting, interactive services, news gathering and other broadband
	satellite applications,'' Oct. 2014.
	
	\bibitem{abramowitz1965handbook}
	M.~Abramowitz and I.~A. Stegun, \emph{Handbook of mathematical functions: with
		formulas, graphs, and mathematical tables}.\hskip 1em plus 0.5em minus
	0.4em\relax North Chelmsford, MA, USA: Courier Corporation, 1965, vol.~55.
	
\end{thebibliography}
\end{document}